\documentclass{emulateapj}

\usepackage{amsmath}
\usepackage{amssymb}
\usepackage{graphicx}
\usepackage{color}
\usepackage{natbib}
\usepackage{epsfig}

\def\gtaprx {\lower .1ex\hbox{\rlap{\raise .6ex\hbox{\hskip .3ex
	{\ifmmode{\scriptscriptstyle >}\else
		{$\scriptscriptstyle >$}\fi}}}
	\kern -.4ex{\ifmmode{\scriptscriptstyle \sim}\else
		{$\scriptscriptstyle\sim$}\fi}}}
\def\ltaprx {\lower .1ex\hbox{\rlap{\raise .6ex\hbox{\hskip .3ex
	{\ifmmode{\scriptscriptstyle <}\else
		{$\scriptscriptstyle <$}\fi}}}
	\kern -.4ex{\ifmmode{\scriptscriptstyle \sim}\else
		{$\scriptscriptstyle\sim$}\fi}}}

\newcommand{\cutt}[1]{\textcolor{blue}{}}

\newcommand{\Ms}{{\ensuremath{{M}_{\odot} }}}
\newcommand{\Zs}{\ensuremath{Z_\odot}}

\newcommand{\Ni}{{\ensuremath{^{56}\mathrm{Ni}}}}

\newcommand{\HII}{{\ion{H}{2}}}

\begin{document}

\title{Population III Hypernovae}

\author{Joseph Smidt\altaffilmark{1}, Daniel J. Whalen\altaffilmark{2}, Brandon K.
Wiggins\altaffilmark{3}, Wesley Even\altaffilmark{4}, Jarrett L. Johnson\altaffilmark{5} 
and Chris L. Fryer\altaffilmark{4}}

\altaffiltext{1}{T-2, Los Alamos National Laboratory, Los Alamos, NM 87545}

\altaffiltext{2}{Universit\"{a}t Heidelberg, Zentrum f\"{u}r Astronomie, Institut f\"{u}r 
Theoretische Astrophysik, Albert-Ueberle-Str. 2, 69120 Heidelberg, Germany}

\altaffiltext{3}{Department of Physics and Astronomy, Brigham Young University, 
Provo, UT  84602}

\altaffiltext{4}{CCS-2, Los Alamos National Laboratory, Los Alamos, NM 87545}

\altaffiltext{5}{XTD-PRI, Los Alamos National Laboratory, Los Alamos, NM 87545}

\begin{abstract}

Population III supernovae have been of growing interest of late for their potential
to directly probe the properties of the first stars, particularly the most energetic 
events that are visible near the edge of the observable universe.  But until now,
hypernovae, the unusually energetic Type Ib/c supernovae that are sometimes
associated with gamma-ray bursts, have been overlooked as cosmic beacons
at the highest redshifts.  In this, the latest of a series of studies on Population III 
supernovae, we present numerical simulations of 25 - 50 \Ms\ hypernovae and
their light curves done with the Los Alamos RAGE and SPECTRUM codes.  We 
find that they will be visible at $z =$ 10 - 15 to the {\it James Webb Space 
Telescope} ({\it JWST}) and $z =$ 4 - 5 to the Wide-Field Infrared Survey 
Telescope (WFIRST), tracing star formation rates in the first galaxies and at the
end of cosmological reionization.  If, however, the hypernova crashes into a 
dense shell ejected by its progenitor it is expected that a superluminous event will 
occur that may be seen at $z \sim$ 20, in the first generation of stars.

\vspace{0.1in}

\end{abstract}


\section{Introduction}

Population III (Pop III) stars ended the cosmic Dark Ages and began cosmological
reionization \citep[e.g.,][]{wan04,oet05,wet08b,wet10} and the chemical enrichment 
of the IGM \citep{mbh03,ss07,bsmith09,ritt12,ss13}.  They also populated the first 
galaxies \citep{jlj09,get10,jeon11,pmb11,wise12,pmb12} and may be the origin of 
supermassive black holes \citep[e.g.,][]{milos09,awa09,th09,pm11,jlj12a,agarw12,
jet13,latif13c,latif13a,jet14}.  Although they are very luminous, individual Pop III 
stars will not be visible to the \textit{James Webb Space Telescope} \citep[\textit{
JWST},][]{jwst06}, the Wide-Field Infrared Survey Telescope (WFIRST), or the 
Thirty-Meter Telescope \citep[TMT; but see][about detecting the \HII\ regions of 
the first stars]{rz12}.  

The fossil abundance record \citep[the ashes of early supernovae thought to be 
imprinted on ancient metal-poor stars, e.g.,][]{bc05,fet05} suggests that some Pop 
III stars were 15 - 50 \Ms\ \citep{jet09b}.  Numerical simulations of primordial star 
formation \citep{on07,turk09,stacy10,clark11,sm11,get11,hos11,get12,stacy12,
susa13,hir13} suggest that Pop III stars were 20 - 500 \Ms\ \citep[for recent reviews, 
see][]{dw12,glov12}. Together, these studies suggest that both high mass and low 
mass Pop III stars existed in the primeval universe, but they do not otherwise 
constrain their properties.  

Primordial SNe \citep[e.g,][]{wet08a} will be the first direct probes of the Pop III initial 
mass function (IMF) because they can be seen at great distances and the masses 
of their progenitors can be inferred from their light curves.  Recent studies have 
shown that Pop III pair-instability (PI) SNe \citep{hw02,fwf10,jw11,kasen11,pan12a,
pan12b,hum12,cw12,cwc13,chen14c} will be visible at $z \ga$ 30 to deep-field 
surveys by \textit{JWST} and at $z \sim 15 - 20$ in all-sky near infrared (NIR) surveys 
by WFIRST and the Wide-Field Imaging Surveyor for High Redshift (WISH) \citep{
wet12b,wet12d,wet12a,wet13d,ds13,ds14,chen14a,smidt14a} \citep[see also][]{jet13a,
wet13a,wet13b,chen14b}.  PI SN candidates have now been identified at low redshifts 
\citep{gy09,cooke12} \citep[see also][]{wet13e}.  Others have found that {\it JWST} will 
detect Pop III core-collapse (CC) SNe at $z \sim$ 10 - 20, depending on the type of 
explosion \citep{wet12e,wet12c,wet13c} \citep[see also][]{tomin11,moriya12,tet12,tet13}.

In the past decade, hypernovae (HNe), with energies that are intermediate to those of 
CC and PI SNe, have been proposed to explain the elemental patterns found in hyper 
metal-poor stars \citep[e.g.,][]{maeda03,Iwamoto2005,Tominaga2007} and to account
for some unusually bright explosions \citep[e.g.,][]{SN1998bw,SN2008D}. Although HNe 
are not fully understood, those observed to date have generally been Type Ib/c SNe and 
have been associated with gamma-ray bursts \citep[GRBs,][]{iwam98,naka01}.  They 
may therefore be explosions of massive stars that have shed their H envelopes and 
are bare He cores.  Since Pop III stars are not generally thought to undergo pulsations 
\citep{baraffe01} or have strong winds \citep{vink01}, the H layer is likely ejected during 
a common envelope phase with a binary companion.  The central engine may be a black 
hole accretion disk system that drives a strong wind or jet that deposits part of its energy 
into the surrounding layers of the star.  The result is a powerful, highly asymmetric 
explosion that can synthesize large amounts of \Ni, both of which may account for its 
brightness.  Because their energies typically range from 10 - 50 $\times$ 10$^{51}$ erg, 
HNe may be visible at redshifts intermediate to those at which PI and CC SNe can be 
detected.  As such, they may be complementary probes of stellar populations in the 
primordial universe. 

We have now calculated light curves and spectra for 25 - 50 \Ms\ Pop III HNe with the 
Los Alamos RAGE and SPECTRUM codes.  In Section 2 we describe our grid of 
RAGE models and how we post process them with SPECTRUM to obtain light curves 
and spectra.  In Section 3 we examine blast profiles, and in Section 4 we show NIR 
light curves and detection thresholds in redshift for HNe. In Section 5 we estimate Pop 
III HN event rates as a function of redshift, and we conclude in Section 6.

\section{Numerical Models}

We calculate light curves and spectra for HNe in three steps. First, stellar collapse 
and explosion is modeled in a 1D Lagrangian hydrodynamics code and its output 
is post processed with an astrophysical nuclear reaction network to obtain 
nucleosynthetic yields.  After explosive burning is complete the blast profiles are 
ported to the RAGE code and evolved out to one year.  We then post process our 
RAGE profiles with the SPECTRUM code to construct light curves and spectra.

\subsection{Collapse and Explosion}

To model collapse and explosion, we use the one-dimensional (1D) Lagrangian 
code and techniques described in \citet{yf07}.  This code includes three-flavor 
neutrino transport with flux-limited diffusion and a coupled set of equations of 
state (EOS) to model the wide range of densities in the collapse phase 
\citep[for details, see][]{het94,fet99}.  It includes a 14-element nuclear network 
\citep{bet89} to follow energy generation.  After collapse, bounce and formation 
of a proto-neutron star, we halt the run and remove the neutron star.  To trigger 
the explosion we inject thermal energy into the innermost 15 zones (roughly 
0.035 \Ms).  Convection mixes this energy fairly uniformly over the convective 
zone.Ê We use 15 zones because they enclose the inner 0.1 \Ms, which is roughly 
the mass of the convection zone.Ê We have verified that varying the number of
convection zones and enclosed mass (0.05 - 0.2 \Ms) yields similar results.  

The duration and magnitude of the energy injection in these artificial explosions 
were adjusted to vary the explosion energies.  During energy injection, the 
protoÐneutron star is modeled as a hard surface.  We do not include neutrino 
flux from the protoÐneutron star, but the energy injected by this flux is minimal 
compared to our artificial energy injection.  Shortly after the end of the energy 
injection we change the hard neutron star surface to an absorbing boundary 
layer to capture the accretion of infalling matter due to neutrino cooling onto the 
protoÐneutron star.  In this manner we can model the explosion out to late times, 
even if there is considerable fallback.

For more accurate yields, we post process our explosions with the public version 
of the torch 
code\footnote{http://cococubed.asu.edu/code\_pages/net\_torch.shtml} \citep{
tim99} using the standard 489 isotope network.  We explode a 25 \Ms\ Pop III star 
with energies of 10, 22, and 52 foe (1 foe $=$ 10$^{51}$ erg) and a 50 \Ms\ star
with energies of 10, 22, 52 and 92 foe.  Profiles for these stars are taken from  
\citet{Woosley2002}, and the masses and energies we have chosen bracket those 
inferred for HNe from observations.  We resolve the inner regions of 25 and 50 
\Ms\ stars with 3085 - 3092 zones and 2093 - 2105 zones, respectively.  Because 
HNe are thought to be powered by jets, or perhaps magnetars, our method for 
energy injection is approximate, and could affect nucleosynthetic yields, light 
curves and spectra for these explosions.

\subsection{RAGE}

We evolve the shock out through the surface of the star and into the surrounding 
medium with the Los Alamos RAGE code \citep{rage}.  RAGE is an adaptive mesh 
refinement (AMR) radiation hydrodynamics code with a second-order conservative 
Godunov hydro scheme and grey or multigroup flux-limited diffusion for modeling 
radiating flows in one, two, or three dimensions (1D, 2D, or 3D). RAGE uses atomic 
opacities compiled from the OPLIB 
database\footnote{http://aphysics2/www.t4.lanl.gov/cgi-bin/opacity/tops.pl}\citep{
oplib} and can evolve multimaterial flows with several options of EOS.  The physics 
in our RAGE models is described in \citet{fet12}:  2-temperature (2T) grey 
flux-limited diffusion, multispecies advection, and energy deposition due to the 
radioactive decay of \Ni{} \citep{fet09}.  We include both the self-gravity of the ejecta 
and the gravity due to the neutron star or black hole point mass that is formed in our 
1D Lagrangian code.  

The point mass is initialized with the mass of the remnant plus any additional 
material that fell back onto it before the model was ported to RAGE.  It can continue 
to grow if there is fallback during the RAGE simulation.  Radiative feedback from the
central object during fallback would, to some degree, regulate infall rates and could 
contribute to the luminosity of the explosion after shock breakout, but we neglect it 
in our simulations.  Self-gravity is calculated with a direct solution to Poisson's 
equation on the 1D spherical AMR grid. It is key to obtaining the correct energy and 
luminosity of the shock because the potential energy of the ejecta while it is still 
inside the star is similar to its kinetic and radiation energies \citep{wet12b}.  After 
shock breakout it is far less important but included for completeness.  We evolve 
mass fractions for H, He, C, N, O, Ne, Mg, Si, S, Ar, Ca, Ti, Cr, Fe and Ni.  

\subsubsection{Model Setup}

Since HNe are associated with Type Ib/c supernovae, we assume that the hydrogen
envelope has been stripped from the star prior to the explosion.  We therefore port 
our explosion profiles to RAGE in three stages. First, we map the region from the 
center of the explosion to the shock in our 1D Lagrangian blast profile onto a uniform 
1D spherical mesh in RAGE.  We then map the original profile of the star from the 
radius of the shock to the surface of the He core to the grid. The H layer is discarded, 
and a wind profile is extended from the surface of the He core out to where its density 
falls to that of the \HII\ region of the star, as described below.  The sharp density drop 
at the surface of the He core is mitigated by an $r^{-20}$ bridge to the wind to avoid 
numerical instabilities at shock breakout. 

The root grid has 100,000 uniform zones with a resolution that varies from 6 $\times$ 
10$^5$ cm to 8 $\times$ 10$^6$ cm.  Up to 2 levels of refinement are performed in 
the initial interpolation of the profiles onto the setup grid and then during the simulation.  
We adopt the error estimator of \citet{Lohner1987} as our refinement criterion, which is 
basically the ratio of the second derivative of a chosen quantity to its first derivative at 
the mesh point at which the error is evaluated.  How this criterion is implemented in 
various geometries is discussed in greater detail in \citet{Almgren2010}.  The result is 
a dimensionless, bounded estimator that allows refinement on any variable according 
to preset error indicators.  We allocate 25\% of this grid to the ejecta profile. The initial 
radius of the shock varies with explosion energy but is typically about half the radius of 
the He core.  

We set outflow and reflecting boundary conditions on the fluid and radiation flows at 
the inner boundary of the mesh, respectively; the former allows us to tally fallback to 
the center of the grid and evolve the point mass.  Outflow conditions are set on both 
flows at the outer boundary. When a run is launched, Courant times are initially short 
due to high temperatures, large velocities and small cell sizes.  To speed up the 
simulation and accommodate the expansion of the flow we resize the grid by a factor 
of 2.5 either every 10$^6$ time steps or when the radiation front has crossed 90\% 
of the grid, whichever happens first.  The initial time step on which the new series 
evolves scales roughly as the ratio of the outer radii of the new and old grids.  We 
again apply up to 2 levels of refinement when mapping the explosion to a new grid 
and throughout the run thereafter.  The properties of our HNe are listed in Table 
\ref{tab:T1}.  Note that higher explosion energies yield larger \Ni\ masses because 
the jet burns more of the core all the way to Ni.  There is a chain of reactions that 
lead to \Ni\ up from the lighter elements, not just O and Si burning like in PI SNe, for 
example.

\begin{deluxetable}{cccccc}  
\tabletypesize{\scriptsize}  
\tablecaption{Hypernova Models (masses are in \Ms\label{tab:T1})}
\tablehead{
\colhead{$M_{\star}$} & \colhead{$R$ ($10^{10}$ cm)}&  \colhead{$E$ (10$^{51}$ 
erg)} & \colhead{$M_{\Ni}$}}
\startdata 
25  &  5.3   &  10  &  0.035   \\
25  &  5.3   &  22  &  0.080   \\
25  &  5.3   &  55  &  0.166   \\
50  &  53.7 &  10  &  0.498   \\
50  &  53.7 &  22  &  1.405   \\
50  &  53.7 &  55  &  1.75   \\  
50  &  53.7 &  92  &  2.12      \vspace{0.05in}   
\enddata 
\end{deluxetable}  
\vspace{0.1in}

\subsubsection{Circumstellar Envelope}

Pop III stars are not generally thought to lose much mass over their lifetimes 
because there are no line-driven winds in their metal-free atmospheres \citep{
kudritzki00,ekstr08}. However, they usually do fully ionize 
their halos and drive out most of the gas, later dying in uniform, low-density H 
II regions \citep[$n \sim$ 0.1 - 1 cm$^{-1}$; e.g.,][]{wan04} \citep[see][about 
the possibility of clumpy circumstellar media]{wn08b,wn08a}.  But the processes 
that strip the H layer from Pop III HN progenitors, such as a common envelope 
phase with a binary companion, a He merger with a binary compact remnant 
companion, or instabilities in the star late in its life \citep[e.g.,][]{fw98,zf01,fry06}, 
reset the density profile in the vicinity of the star.  This is true if the star dies in a 
cosmological halo at $z \sim$ 20 or in a protogalaxy at $z \sim$ 10 - 15.

\begin{figure*}
\begin{center}
\begin{tabular}{cc}
\epsfig{file=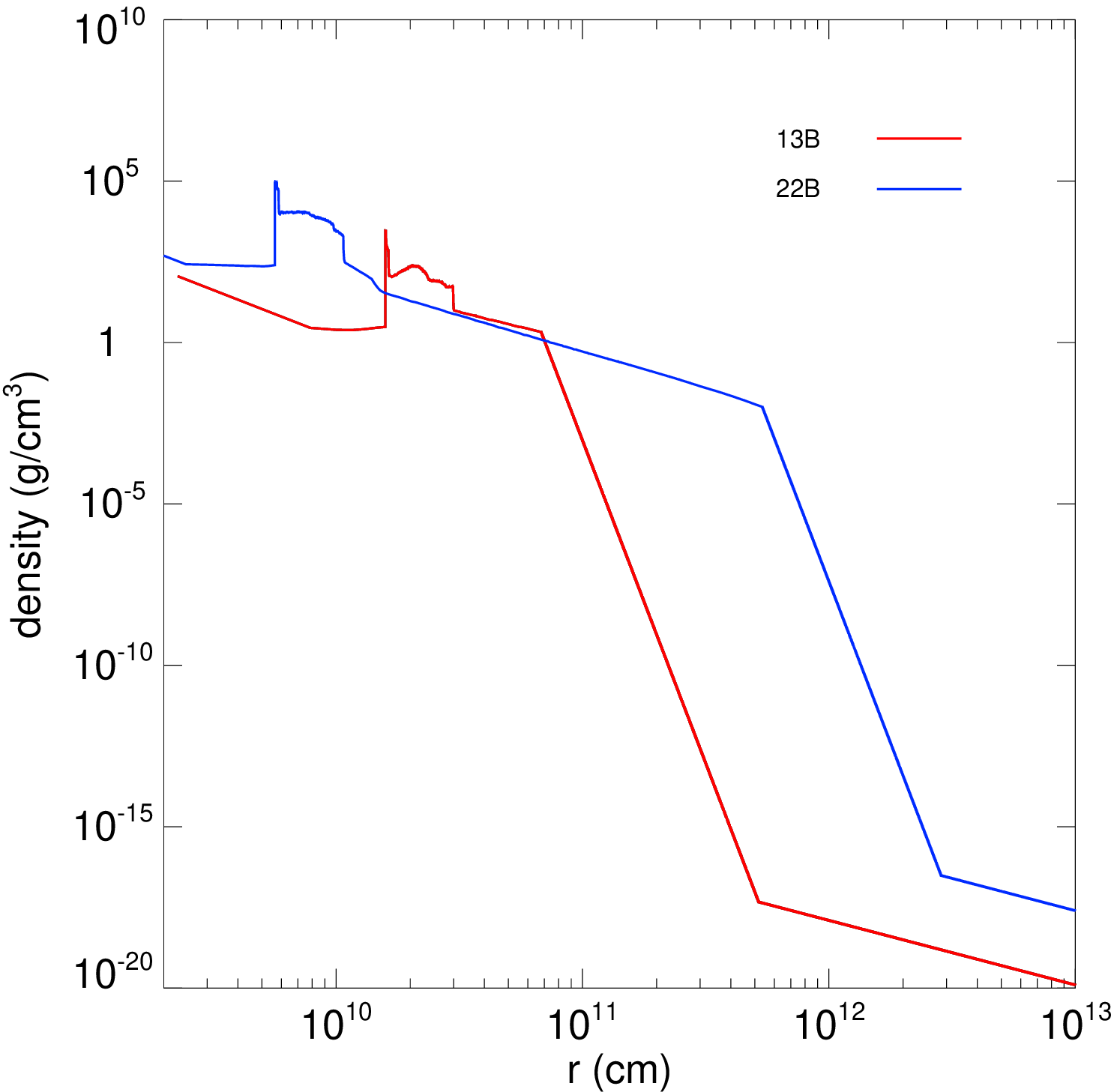,width=0.45\linewidth,clip=} & 
\epsfig{file=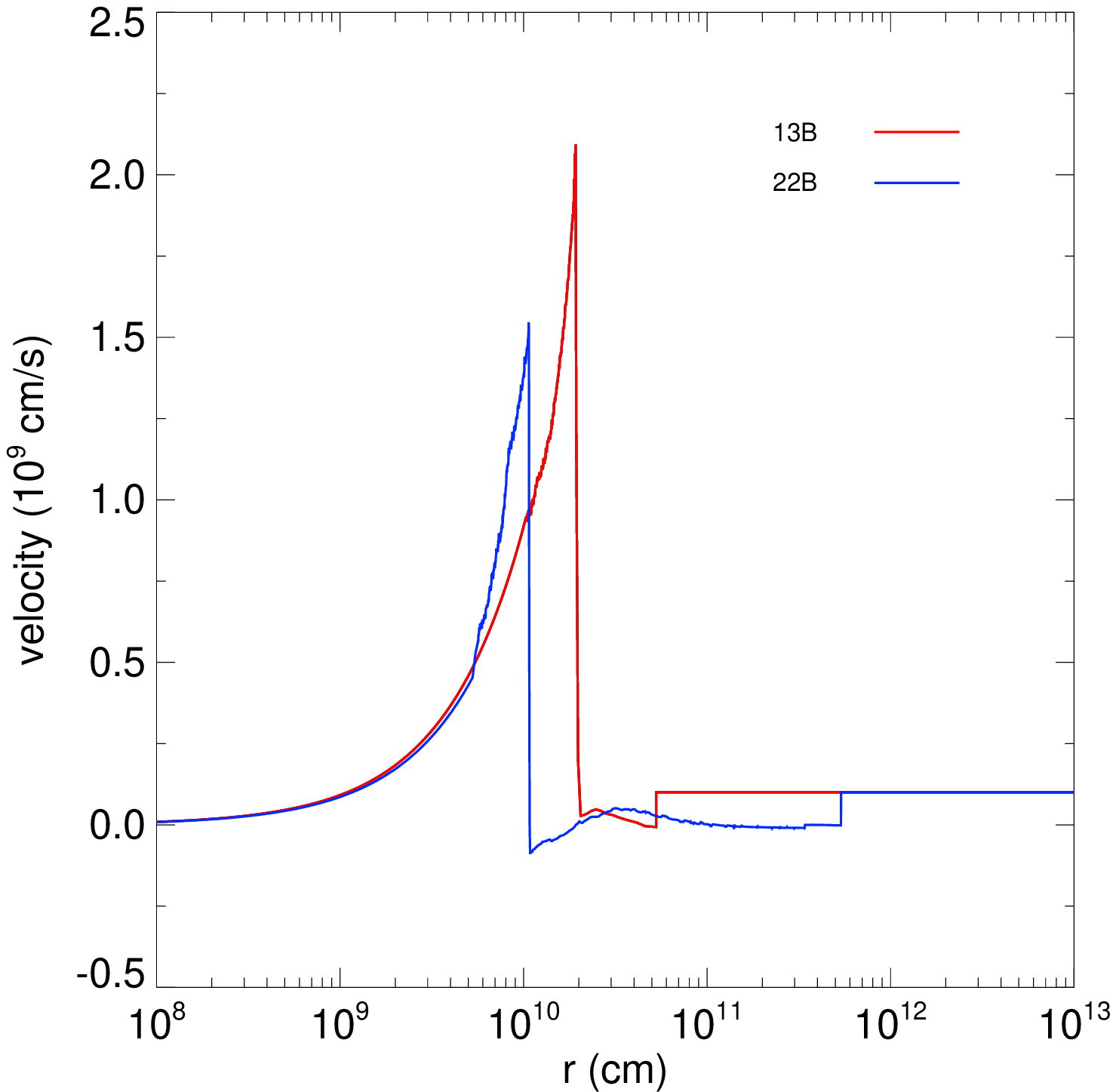,width=0.45\linewidth,clip=} \\
\end{tabular}
\end{center}
\caption{Profiles for the shock, the star and its surrounding envelope as initialized in 
RAGE. Red: 13 foe HN, 25 \Ms\ progenitor. Blue: 22 foe HN, 50 \Ms\ progenitor. Left:  
densities.  Right:  velocities.}
\label{fig:initprof}
\end{figure*}

The expulsion of the envelope usually proceeds as an outburst that ejects a massive 
shell that is followed by a fast wind.  If the shell is less than $\sim$ 0.01 pc from the 
star when it dies, ejecta from the HN will crash into it and make a superluminous Type 
IIn SN \citep[e.g.,][]{moriya10,moriya12,wet12e}.  For simplicity, we assume that the 
shell has been driven beyond 1 pc so there is no collision and it is too diffuse to 
attenuate light from the explosion.  We thus join a simple low-mass wind profile to the 
surface of the star:
\vspace{0.05in}
\begin{equation}
\rho_\mathrm{w}(r) = \frac{\dot{m}}{4 \pi r^2 v_\mathrm{w}}, \vspace{0.05in}
\end{equation}
where $\dot{m}$ is the mass loss rate of the wind and $v_\mathrm{w}$ is its speed.  
We take $v_\mathrm{w}$ to be 1000 km s$^{-1}$ and the H and He mass fractions 
in the wind to be 76\% and 24\%, respectively.  We choose $\dot{m}$ to yield $\rho_
\mathrm{w} \sim$ 2 $\times$ 10$^{-18}$ g cm$^{-3}$ at the bottom of the density 
bridge from the surface of the star.  This choice of wind ensures that it is optically 
thin at the bottom of the bridge but still dense enough to prevent numerical 
instabilities in the radiation solution there.  The wind profile is continued along the 
grid until its density falls to $n = $ 0.1 cm$^{-3}$, that of the \HII\ region of the star.  
The wind is then replaced by this uniform \HII\ region.  We show some initial density 
and velocity profiles for our RAGE models in Figure~\ref{fig:initprof}.

\subsection{SPECTRUM} 

To calculate spectra from a RAGE blast profile we map its densities, temperatures, 
velocities and mass fractions onto a 2D grid in the Los Alamos SPECTRUM code.  
SPECTRUM then performs a direct sum of the luminosity of every fluid element in 
the discretized profile to compute the total flux escaping the ejecta along the line of 
sight at every wavelength.  This procedure accounts for Doppler shifts and time 
dilation due to the relativistic expansion of the ejecta.  SPECTRUM also calculates 
intensities of emission lines and the attenuation of flux along the line of sight, 
capturing both limb darkening and absorption lines imprinted on the flux by 
intervening material in the ejecta and wind.  Each spectrum has 14899 wavelengths.

We first extract gas densities, velocities, mass fractions and radiation temperatures 
from the AMR hierarchy in RAGE and order them by radius.  Because of constraints 
on machine memory and time, only a subset of these points are used in SPECTRUM.  
We determine the position of the radiation front, which is taken to be where $aT^4$ 
rises above 10$^{-4}$ erg/cm$^3$.  Next, we find the radius of the $\tau = $ 40 
surface by integrating the optical depth due to Thomson scattering in from the outer 
boundary, taking $\kappa_{Th}$ to be 0.288 for H and He gas at the mass fractions 
in the wind \citep[see Section 2.4 of][]{wet12c}.  This is the greatest depth from which 
most of the radiation can escape from the ejecta.  

The extracted gas densities, velocities, temperatures and species mass fractions are 
then interpolated onto a 2D grid in $r$ and $\theta$ in SPECTRUM.  The inner mesh
boundary is the same as in RAGE and the outer boundary is 10$^{18}$ cm.  Eight 
hundred uniform zones in log $r$ are assigned from the center of the grid to the $\tau 
=$ 40 surface, and the region from the $\tau =$ 40 surface to the radiation front is 
partitioned into 6200 uniform zones in $r$.  The wind between the front and the outer 
edge of the grid is divided into 500 uniform zones in log $r$, for a total of 7500 radial 
bins.  The variables within each of these new radial bins are mass averaged so that 
the SPECTRUM profile reproduces very sharp features from the RAGE profile.  The 
mesh is uniformly divided into 160 bins in $\mu =$ cos$\,\theta$ from -1 to 1. 

Our grid fully resolves regions of the flow from which photons can escape the ejecta 
and only lightly samples those from which most cannot.  We use a 2D grid in 
SPECTRUM even though our RAGE profiles are only 1D to approximate effects like 
limb darkening and P Cygni profiles. For example, photons approaching an observer 
from the leading edge of the fireball traverse different path lengths through the ejecta 
than those coming from the poles, and mapping to a 2D grid in SPECTRUM partially 
captures these effects on the overall luminosity reaching a distant point.

\section{Blast Profiles}

\begin{figure}
\begin{center}
\begin{tabular}{c}
\epsfig{file=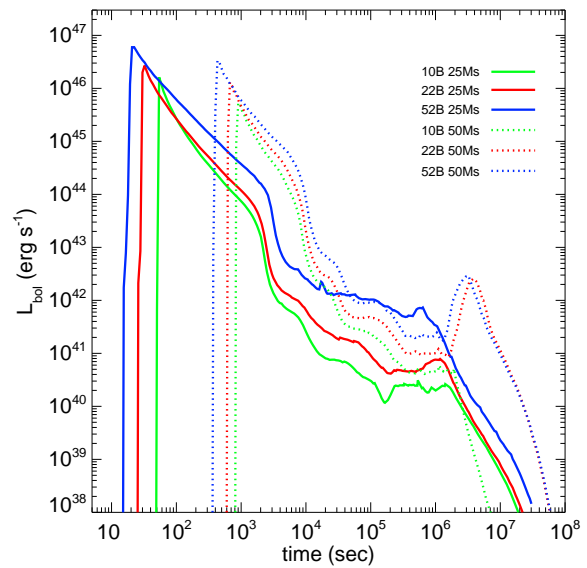,width=0.9\linewidth,clip=} 
\end{tabular}
\end{center}
\caption{Bolometric light curves for all six HNe.}
\label{fig:bolo}
\end{figure}

\begin{figure*}
\begin{center}
\begin{tabular}{ccc}
\epsfig{file=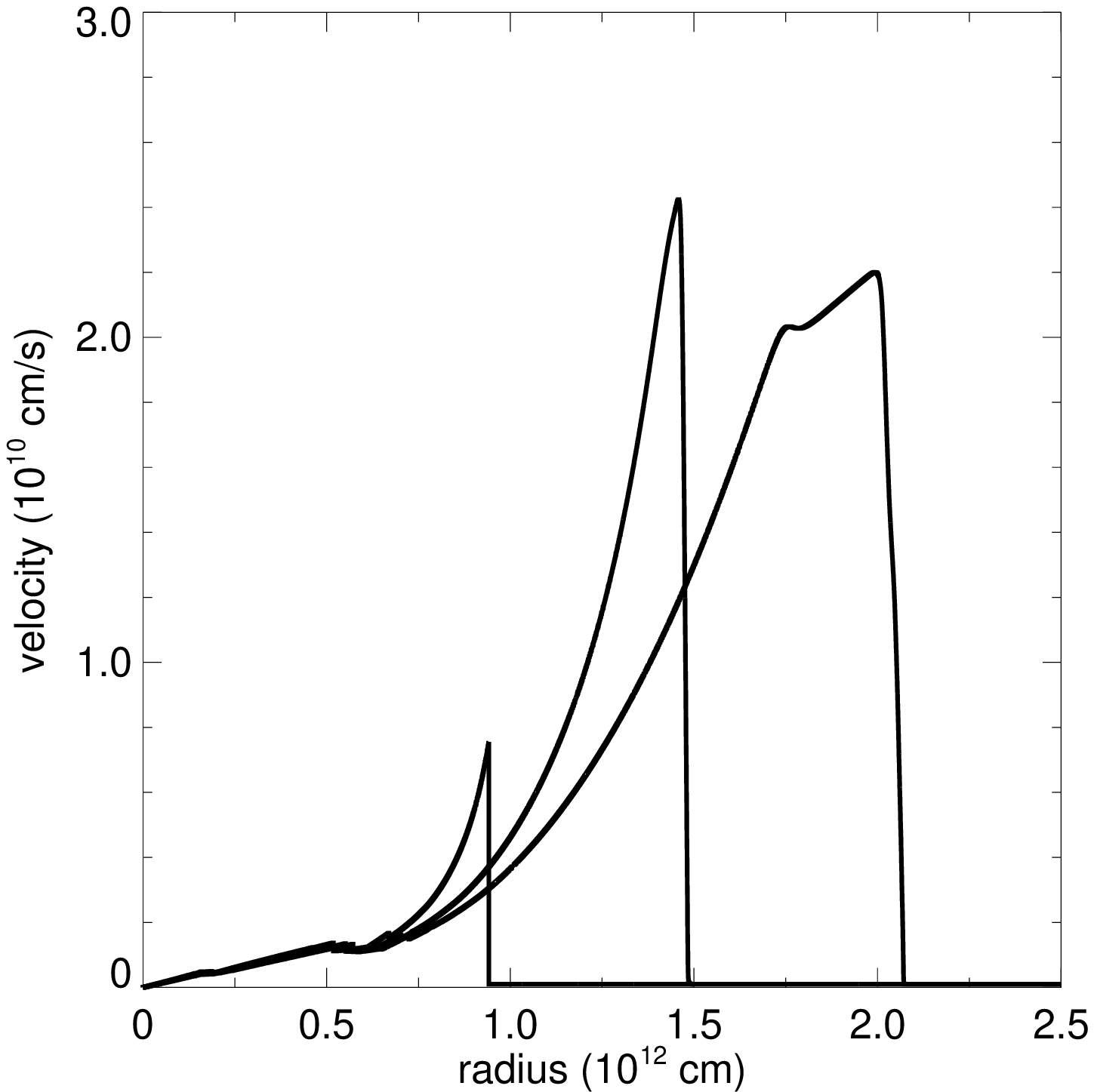,width=0.3\linewidth,clip=}  &
\epsfig{file=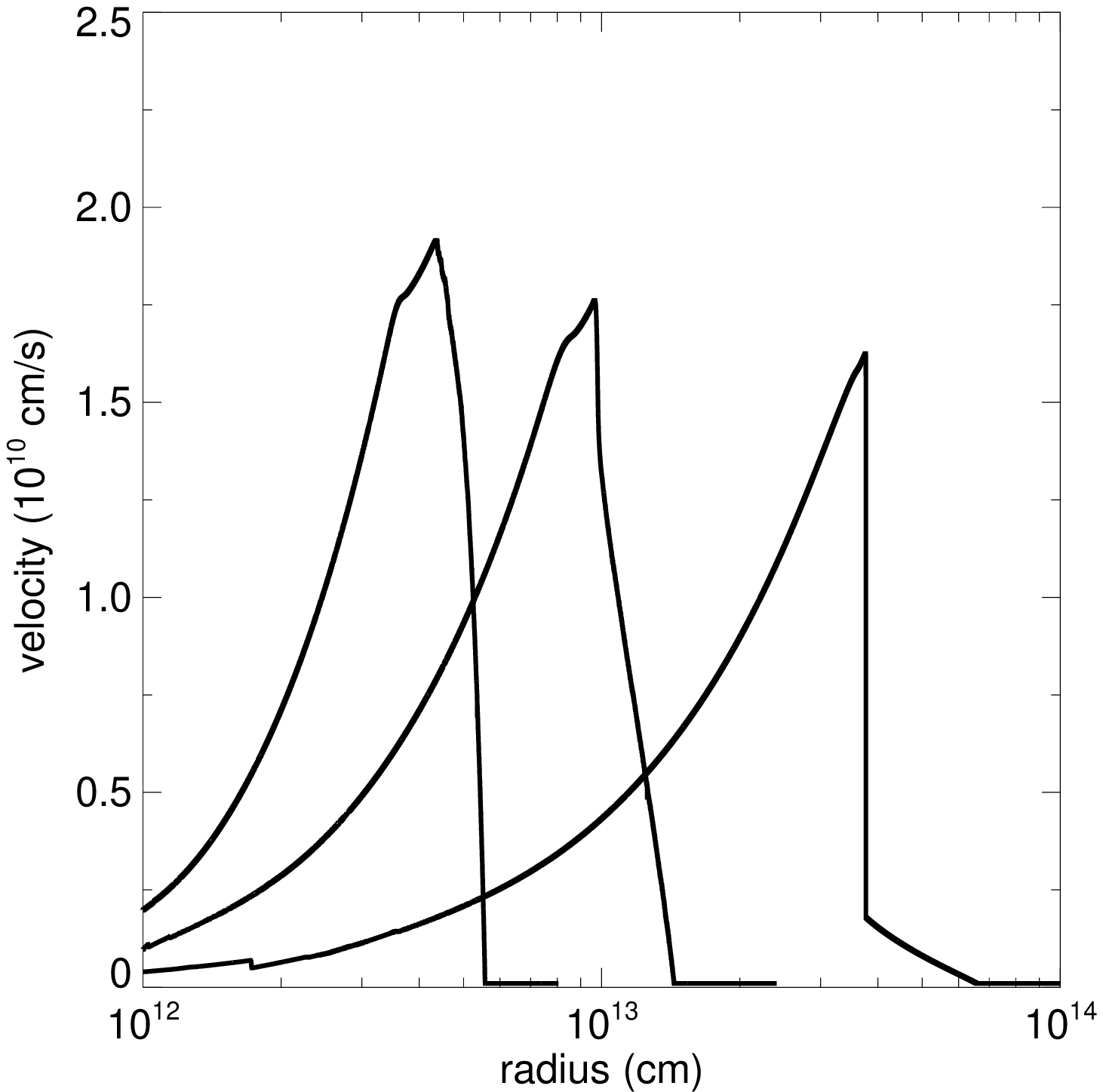,width=0.3\linewidth,clip=}  &
\epsfig{file=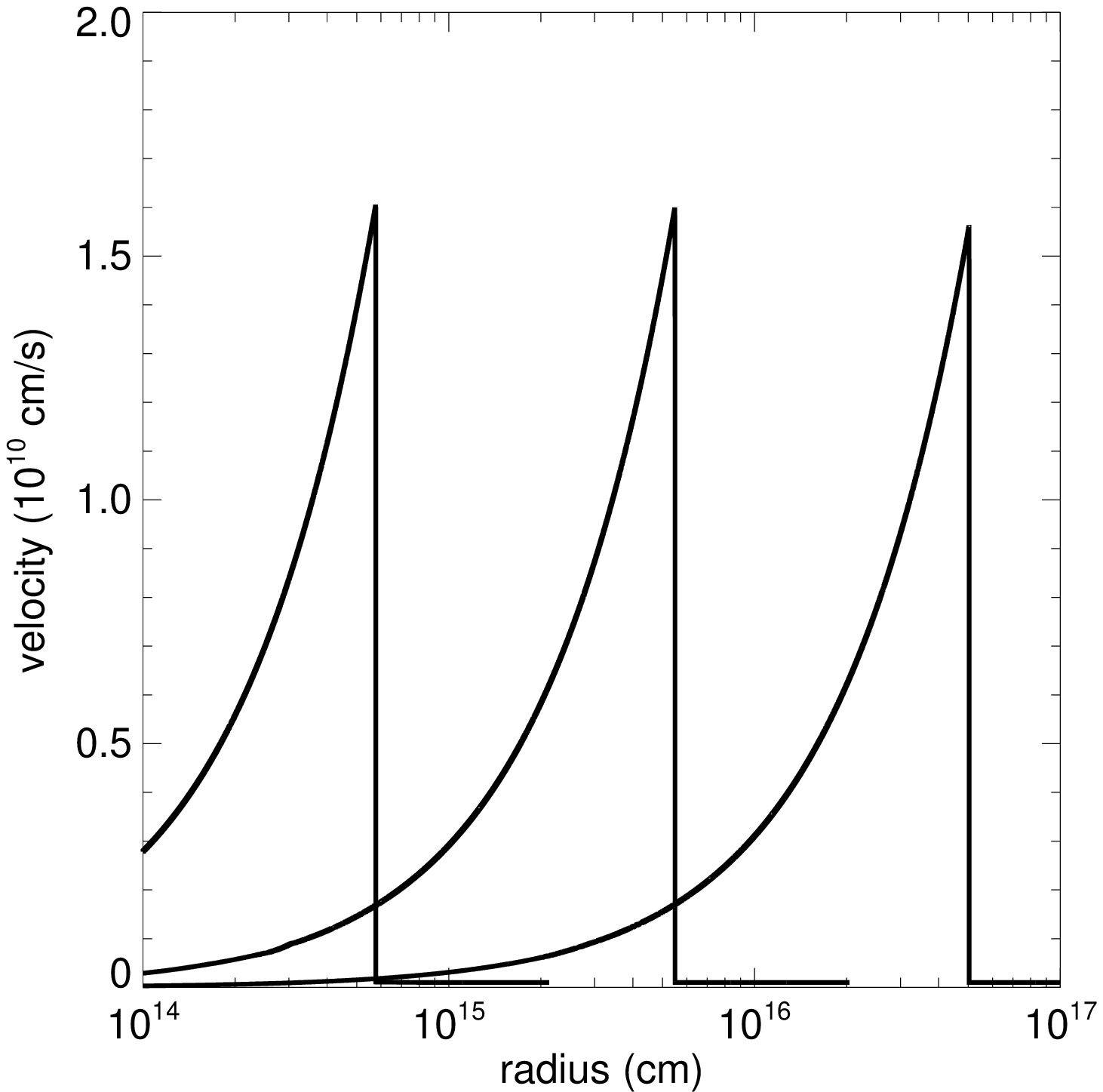,width=0.3\linewidth,clip=}  \\
\epsfig{file=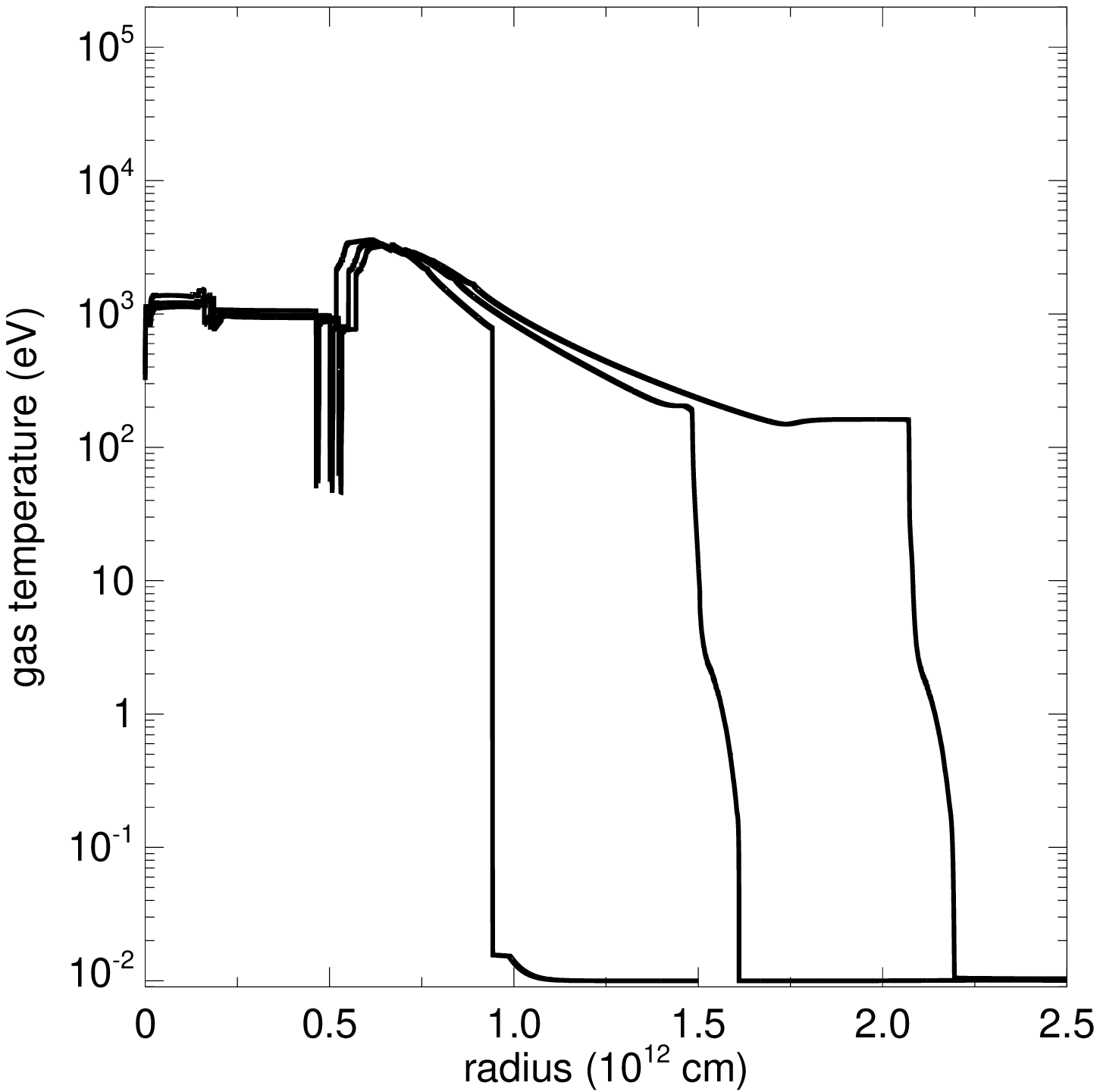,width=0.3\linewidth,clip=}  &
\epsfig{file=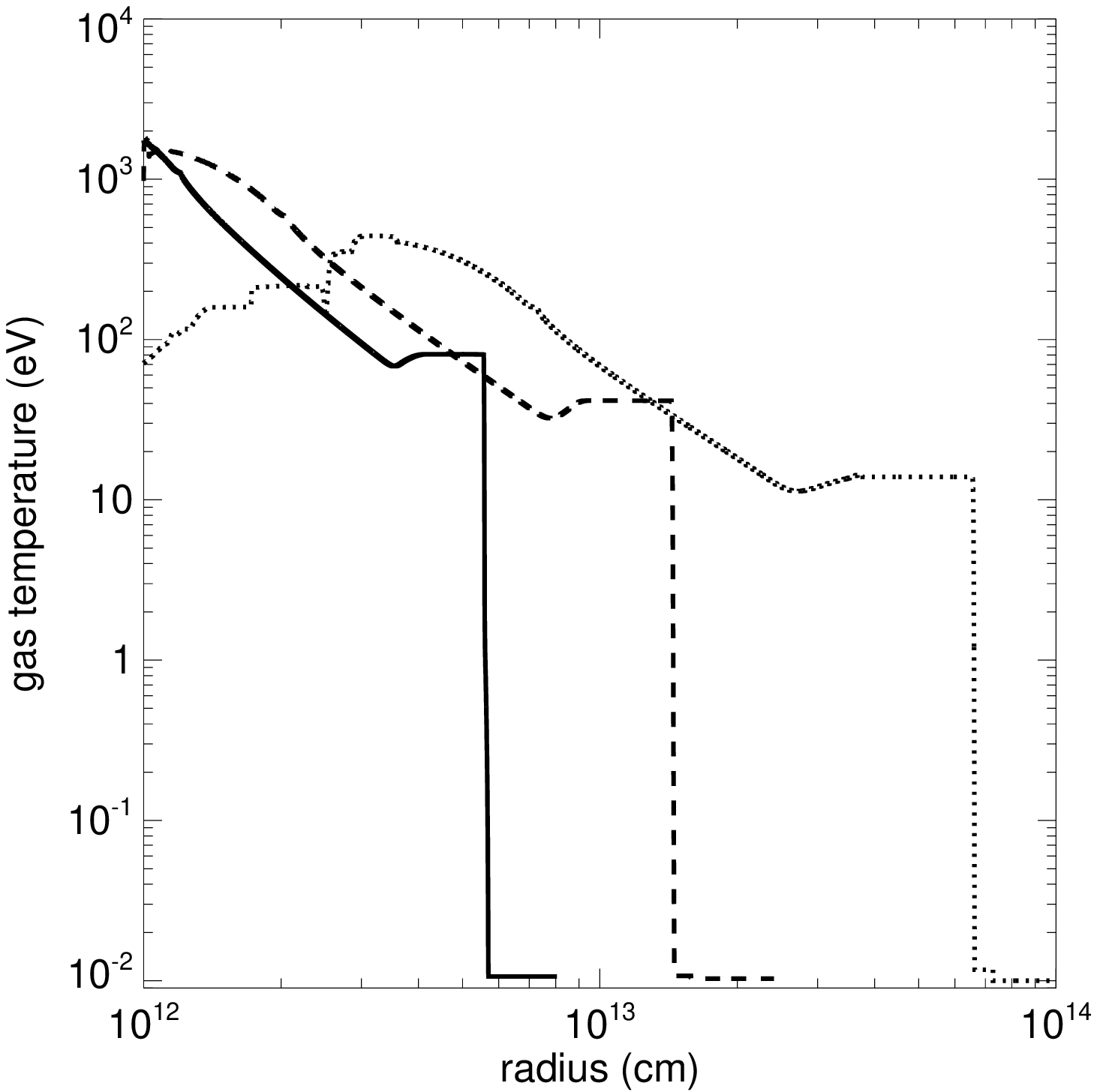,width=0.3\linewidth,clip=}  & 
\epsfig{file=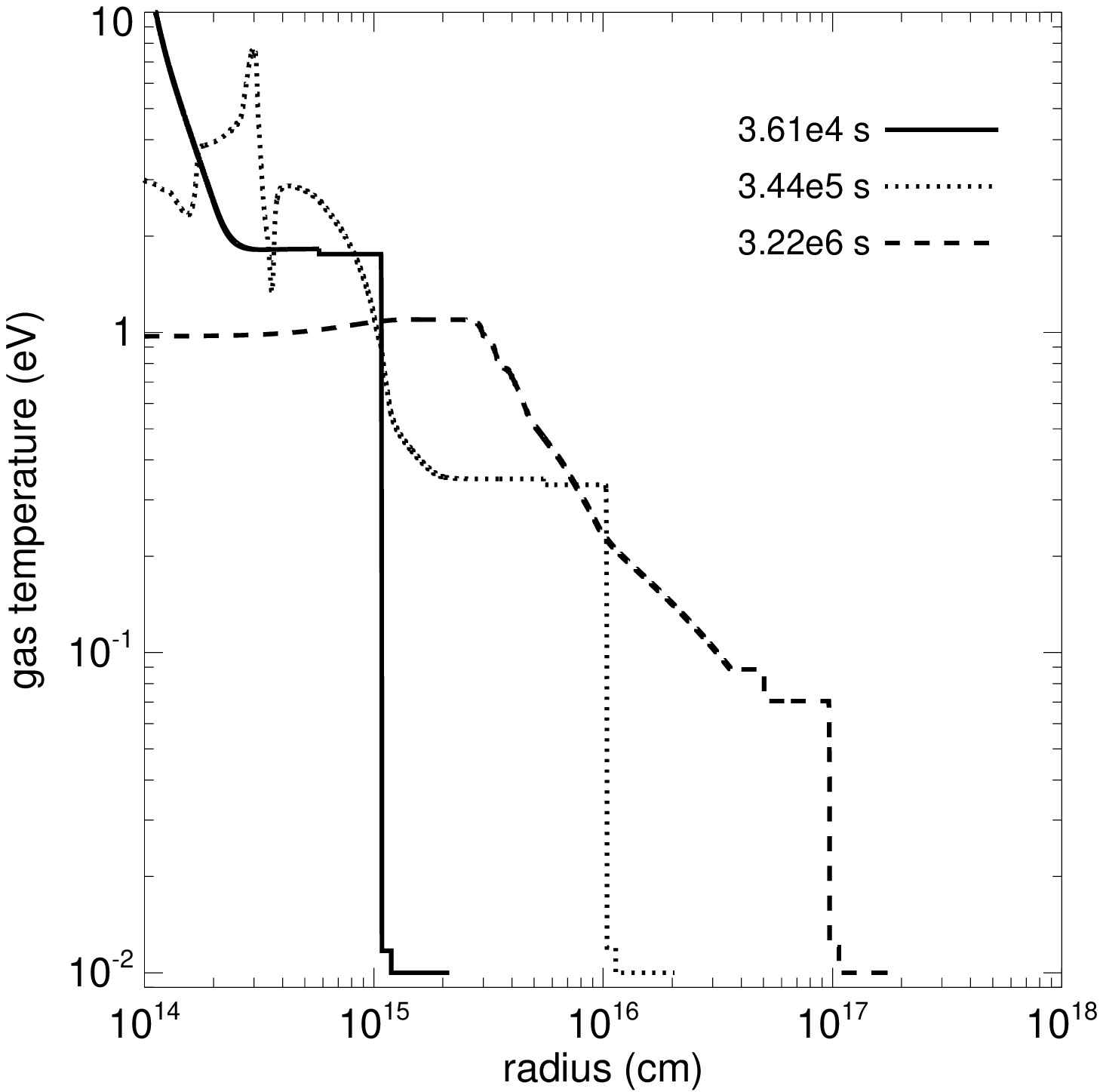,width=0.3\linewidth,clip=}  \\
\epsfig{file=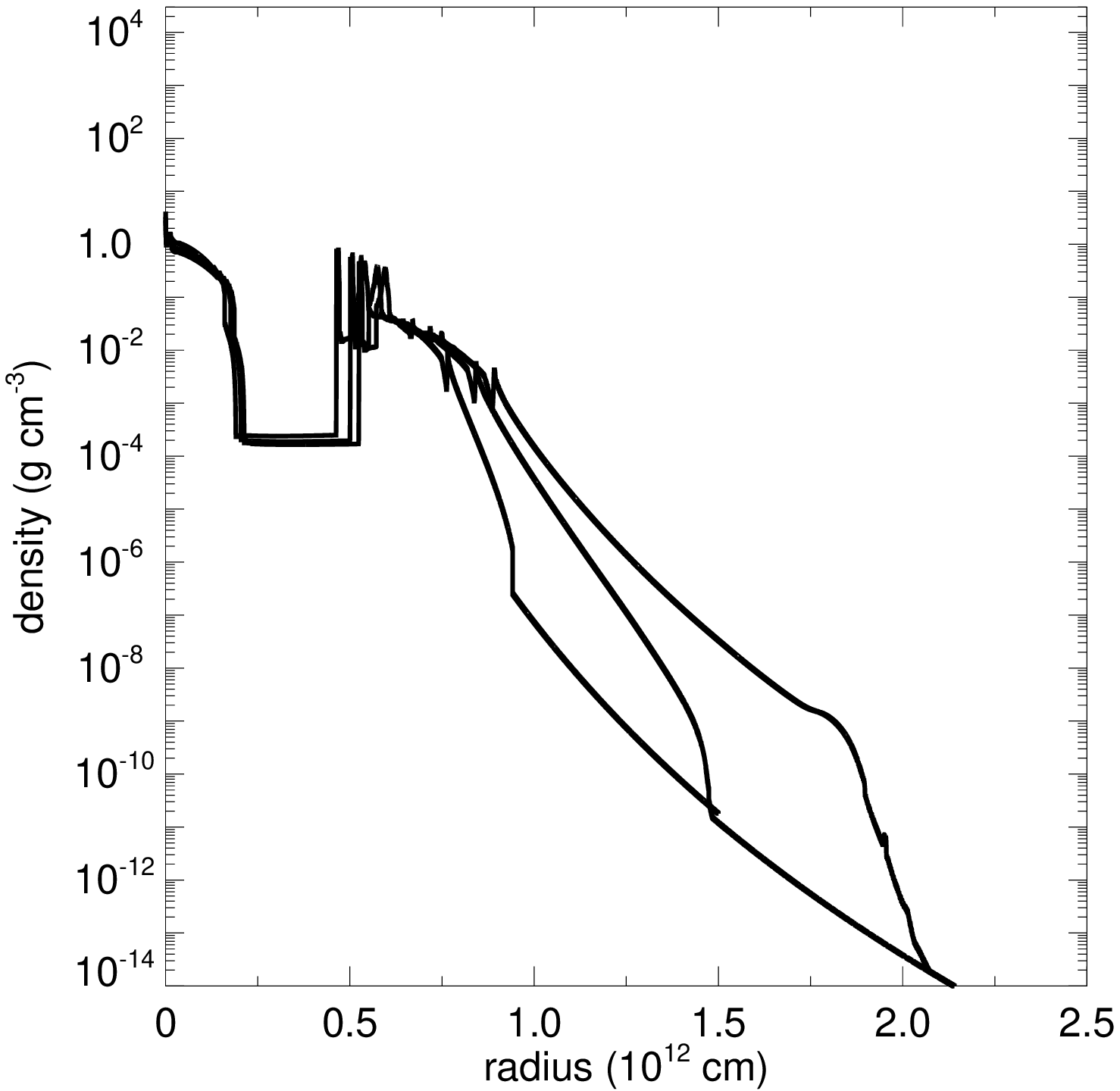,width=0.3\linewidth,clip=}  &
\epsfig{file=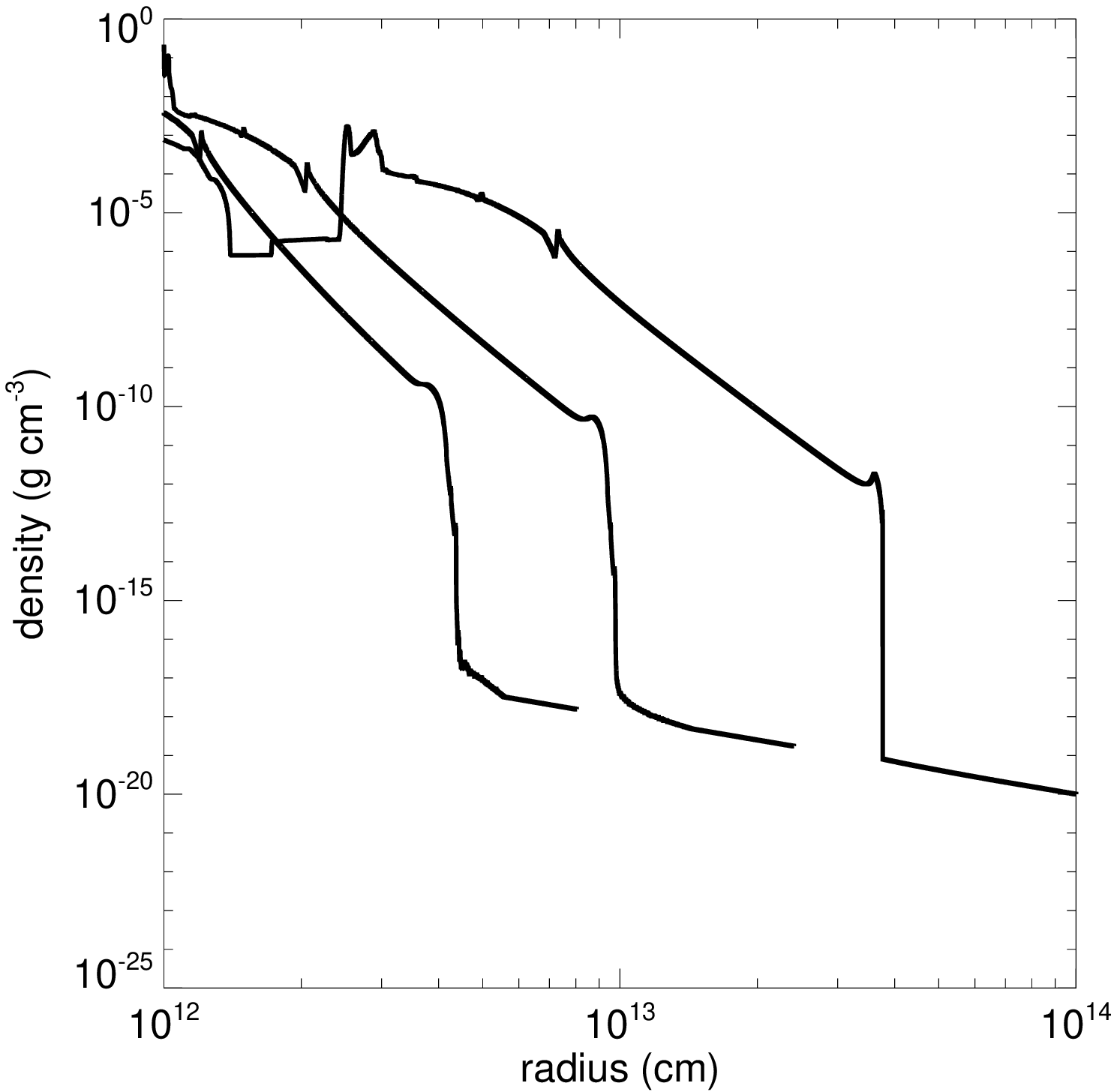,width=0.3\linewidth,clip=}  &
\epsfig{file=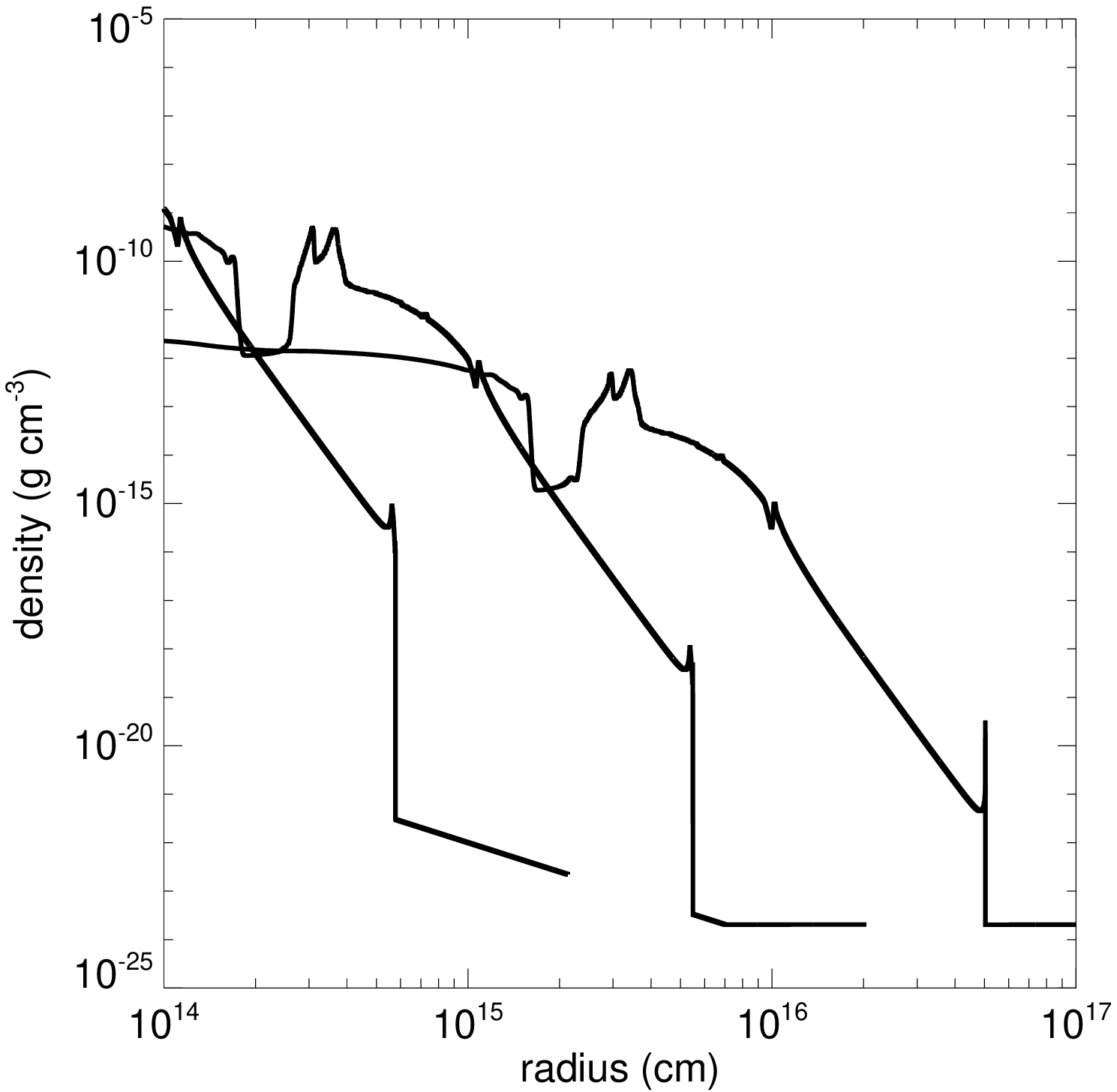,width=0.3\linewidth,clip=}
\end{tabular}
\end{center}
\caption{Hydrodynamic evolution of the 52 foe 50 \Ms\ Hn.  Top:  velocities; center: 
temperatures; bottom:  densities.  Left: shock breakout. From left to right the times 
are 342 s, 373 s, and 392 s. Center:  intermediate evolution.  From left to right, the 
times are 508 seconds, 800 seconds and 2508 seconds. Right:  later evolution.  
From left to right, the times are 3.61 $\times$ 10$^4$ seconds, 3.44 $\times$ 10$^
5$ seconds and 3.22 $\times$ 10$^6$ seconds.}
\label{fig:hydro}
\end{figure*}

\begin{figure*}
\begin{center}
\begin{tabular}{c}
\epsfig{file=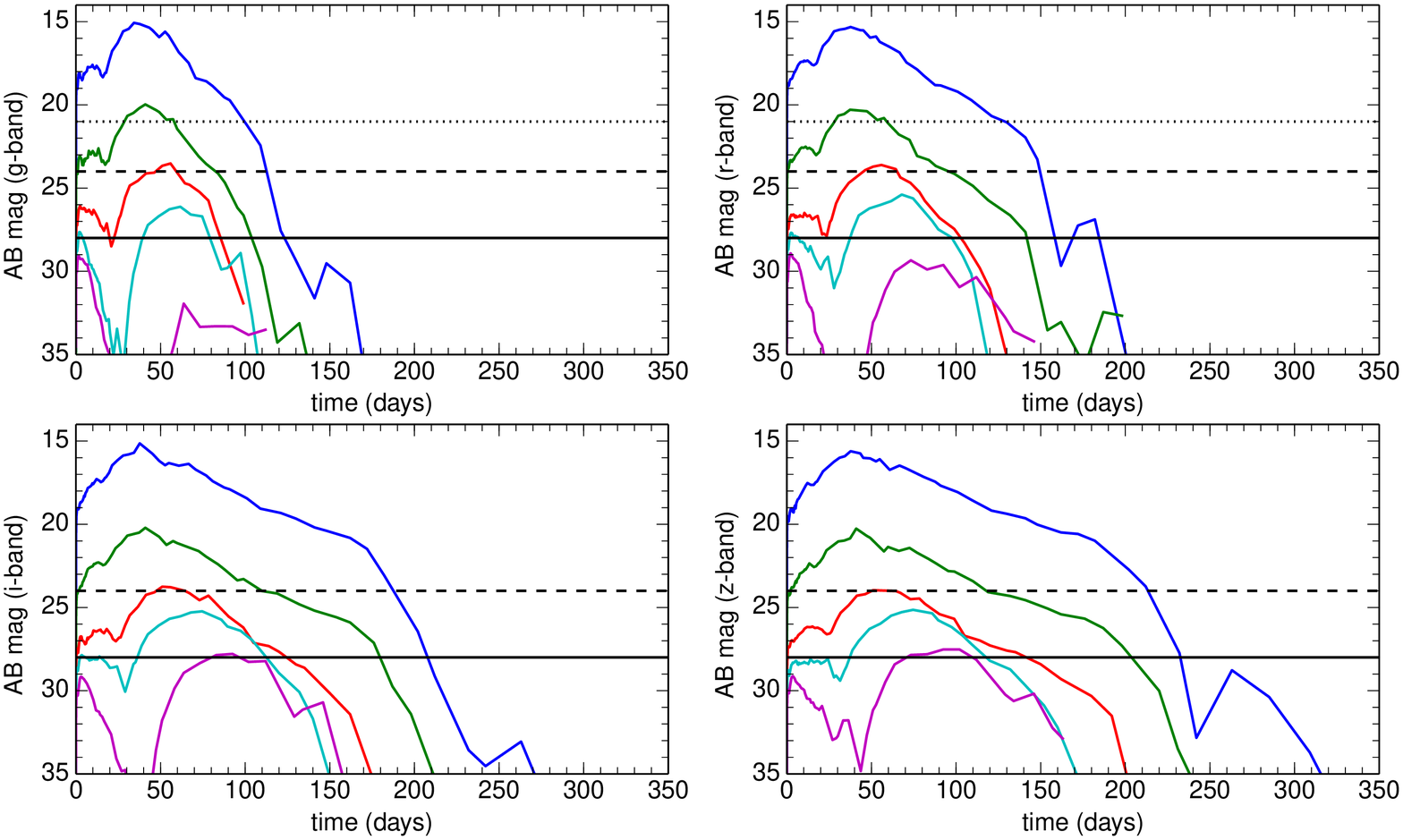,width=0.99\linewidth,clip=} \\
\epsfig{file=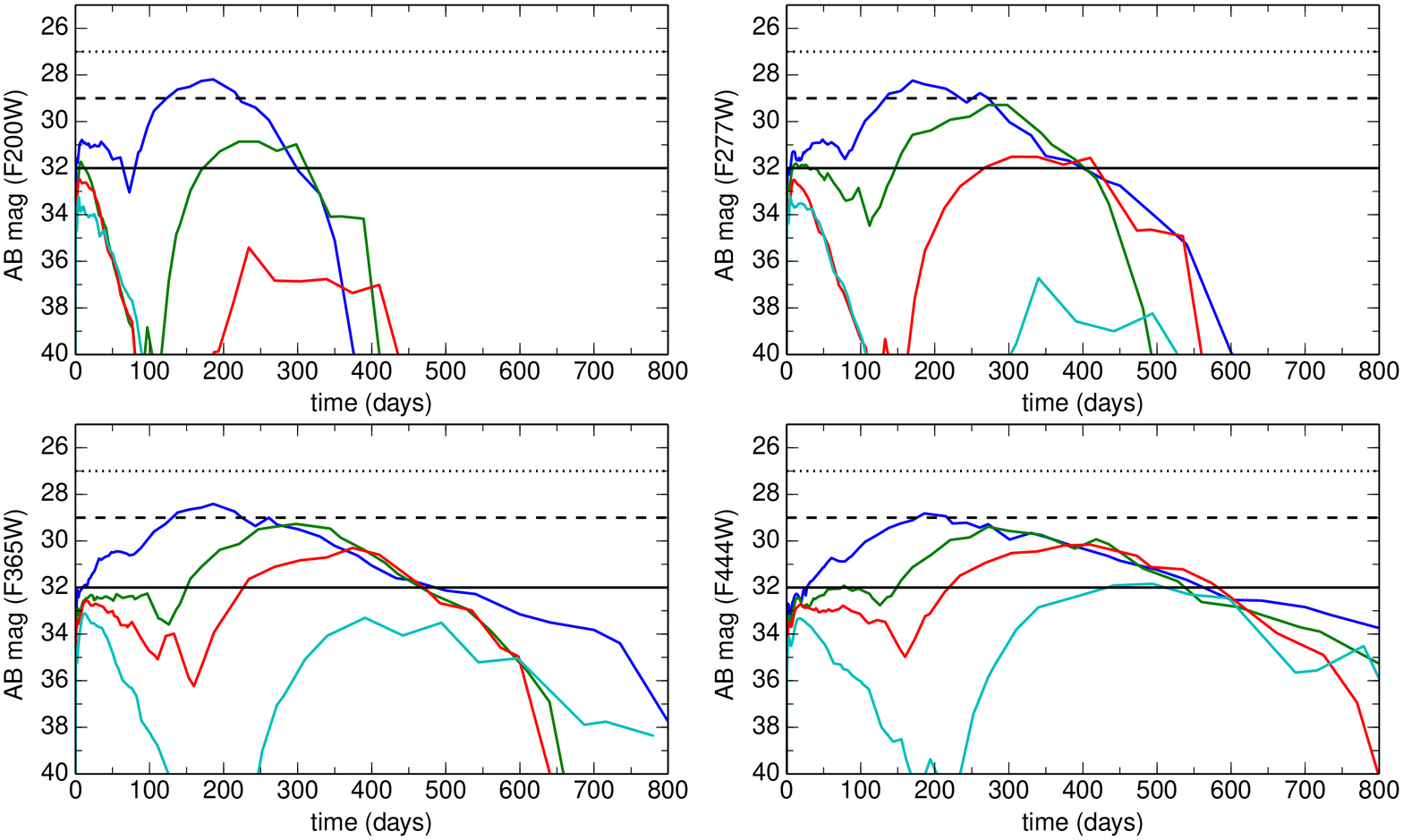,width=0.99\linewidth,clip=}
\end{tabular}
\end{center}
\caption{Light curves for the 52 foe 50 \Ms\ HN at low redshifts (upper panels) and 
high redshifts (upper panels).  In the upper panels, $z =$ 0.01 ({\it dark blue}), 0.1 
({\it green}), 0.5 ({\it red}), 1 ({\it light blue}), and 2 ({\it purple}).  The horizontal 
dotted, dashed and solid lines are detection limits for PTF, Pan-STARRS and LSST, 
respectively. In the lower panels, $z =$ 4 ({\it dark blue}), 7 ({\it green}), 10 ({\it red}), 
15 ({\it light blue}) and 20 ({\it purple}). The horizontal dotted, dashed and solid lines 
are detection limits for WFIRST, WFIRST with spectrum stacking and {\it JWST}, 
respectively.  The wavelength of each filter can be read from its name; for example, 
the F277W filter is centered at 2.77 $\mu$m, and so forth.}
\label{fig:52mag}
\end{figure*}

\begin{figure*}
\begin{center}
\begin{tabular}{c}
\epsfig{file=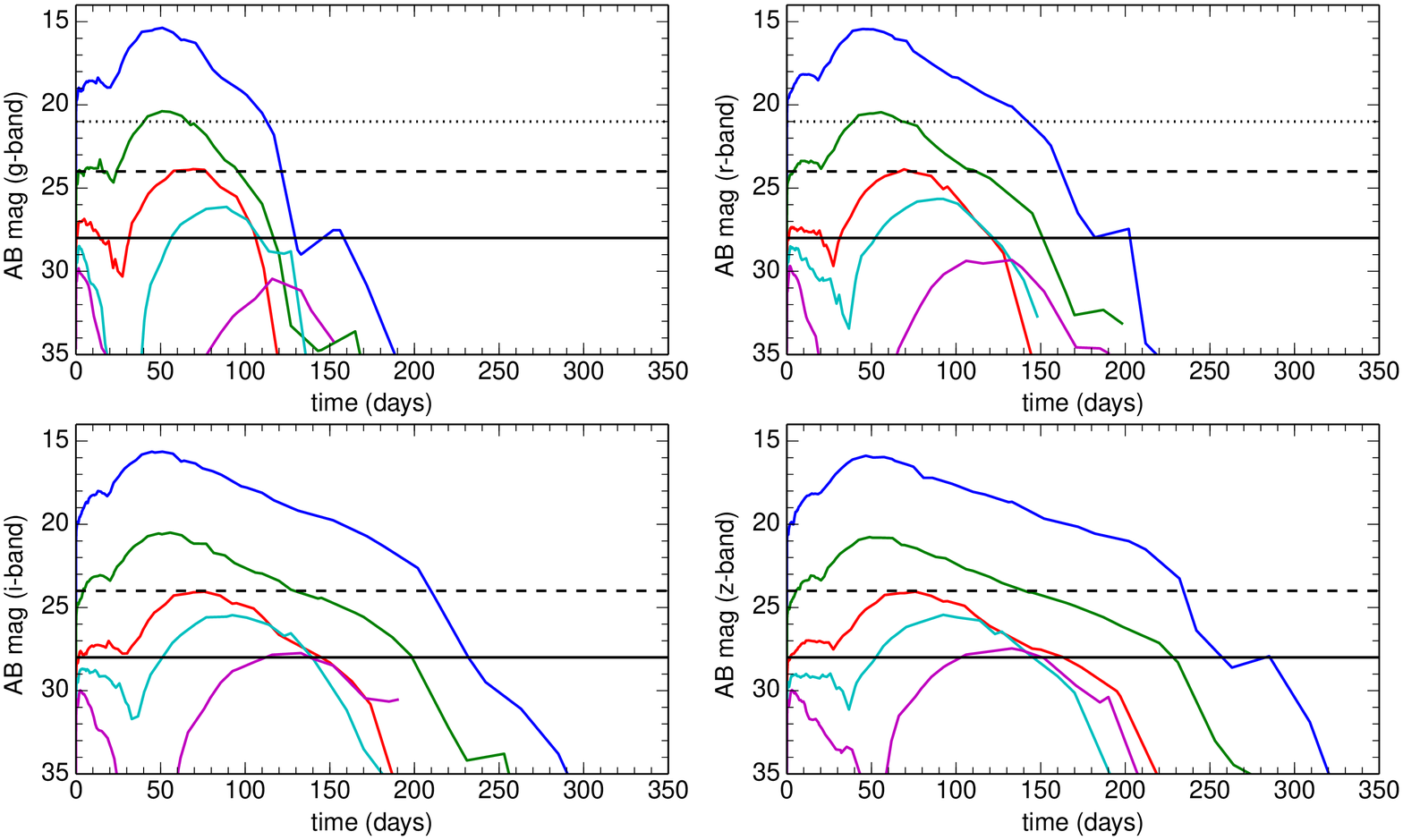,width=0.99\linewidth,clip=} \\
\epsfig{file=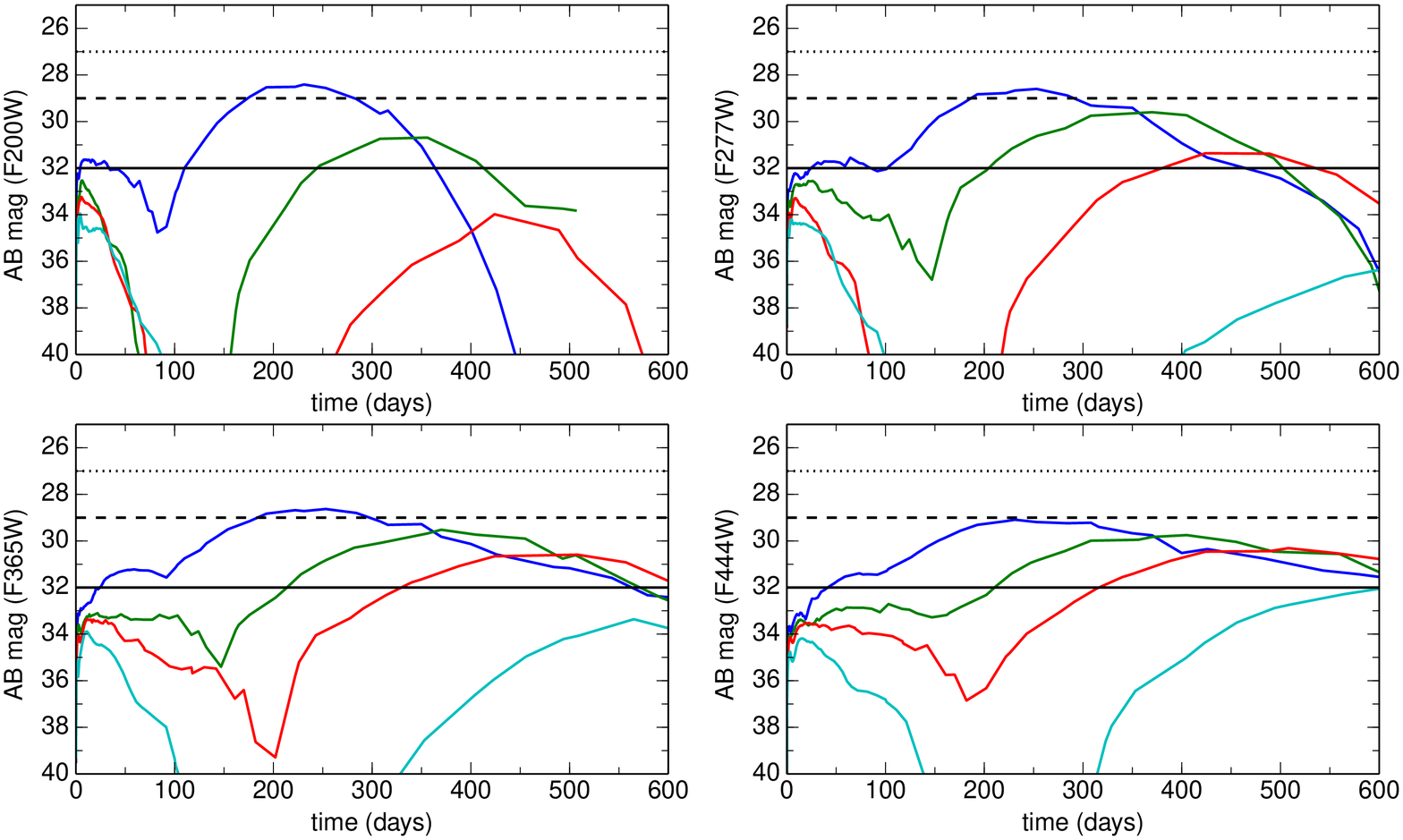,width=0.99\linewidth,clip=}
\end{tabular}
\end{center}
\caption{Light curves for the 22 foe 50 \Ms\ HN at low redshifts (upper panels) and 
high redshifts (upper panels).  In the upper panels, $z =$ 0.01 ({\it dark blue}), 0.1 
({\it green}), 0.5 ({\it red}), 1 ({\it light blue}), and 2 ({\it purple}).  The horizontal 
dotted, dashed and solid lines are detection limits for PTF, Pan-STARRS and LSST, 
respectively. In the lower panels, $z =$ 4 ({\it dark blue}), 7 ({\it green}), 10 ({\it red}), 
15 ({\it light blue}) and 20 ({\it purple}). The horizontal dotted, dashed and solid lines 
are detection limits for WFIRST, WFIRST with spectrum stacking and {\it JWST}, 
respectively.  The wavelength of each filter can be read from its name; for example, 
the F277W filter is centered at 2.77 $\mu$m, and so forth.}
\label{fig:22mag}
\end{figure*}

\begin{figure*}
\begin{center}
\begin{tabular}{c}
\epsfig{file=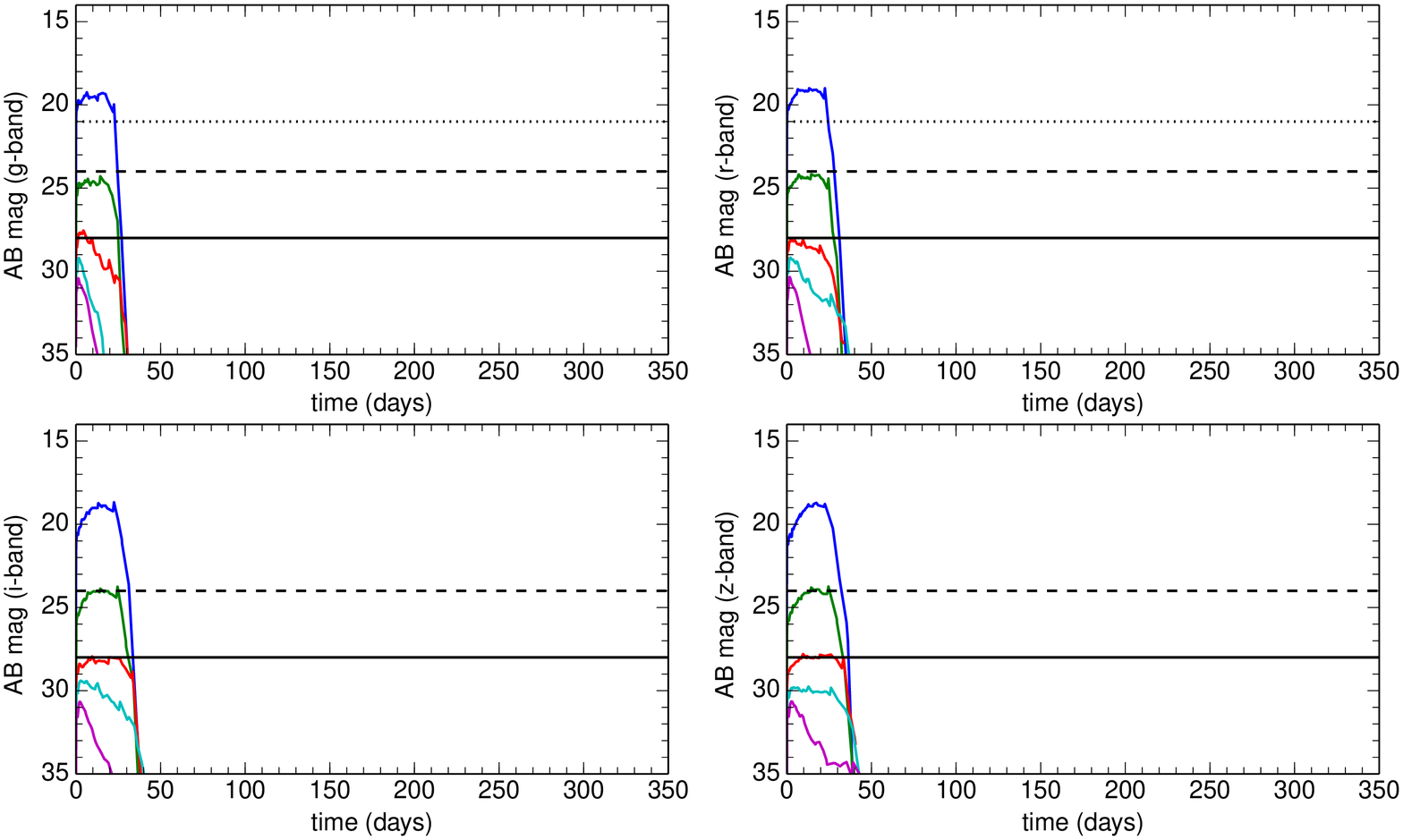,width=0.99\linewidth,clip=} \\
\epsfig{file=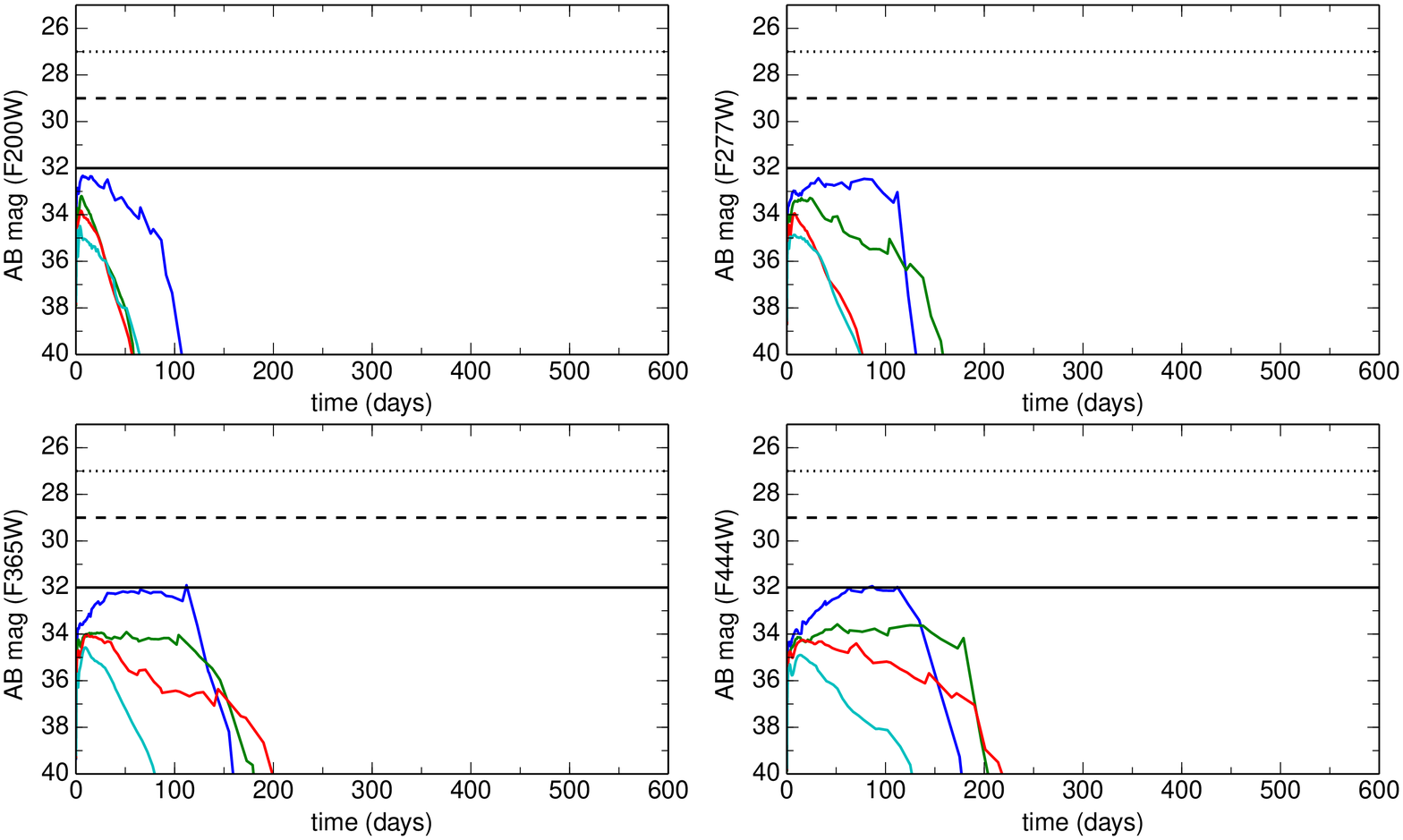,width=0.99\linewidth,clip=}
\end{tabular}
\end{center}
\caption{Light curves for the 10 foe 50 \Ms\ HN at low redshifts (upper panels) and 
high redshifts (upper panels).  In the upper panels, $z =$ 0.01 ({\it dark blue}), 0.1 
({\it green}), 0.5 ({\it red}), 1 ({\it light blue}), and 2 ({\it purple}).  The horizontal 
dotted, dashed and solid lines are detection limits for PTF, Pan-STARRS and LSST, 
respectively. In the lower panels, $z =$ 4 ({\it dark blue}), 7 ({\it green}), 10 ({\it red}), 
15 ({\it light blue}) and 20 ({\it purple}). The horizontal dotted, dashed and solid lines 
are detection limits for WFIRST, WFIRST with spectrum stacking and {\it JWST}, 
respectively.  The wavelength of each filter can be read from its name; for example, 
the F277W filter is centered at 2.77 $\mu$m, and so forth.}
\label{fig:10mag}
\end{figure*}

\begin{figure*}
\begin{center}
\begin{tabular}{c}
\epsfig{file=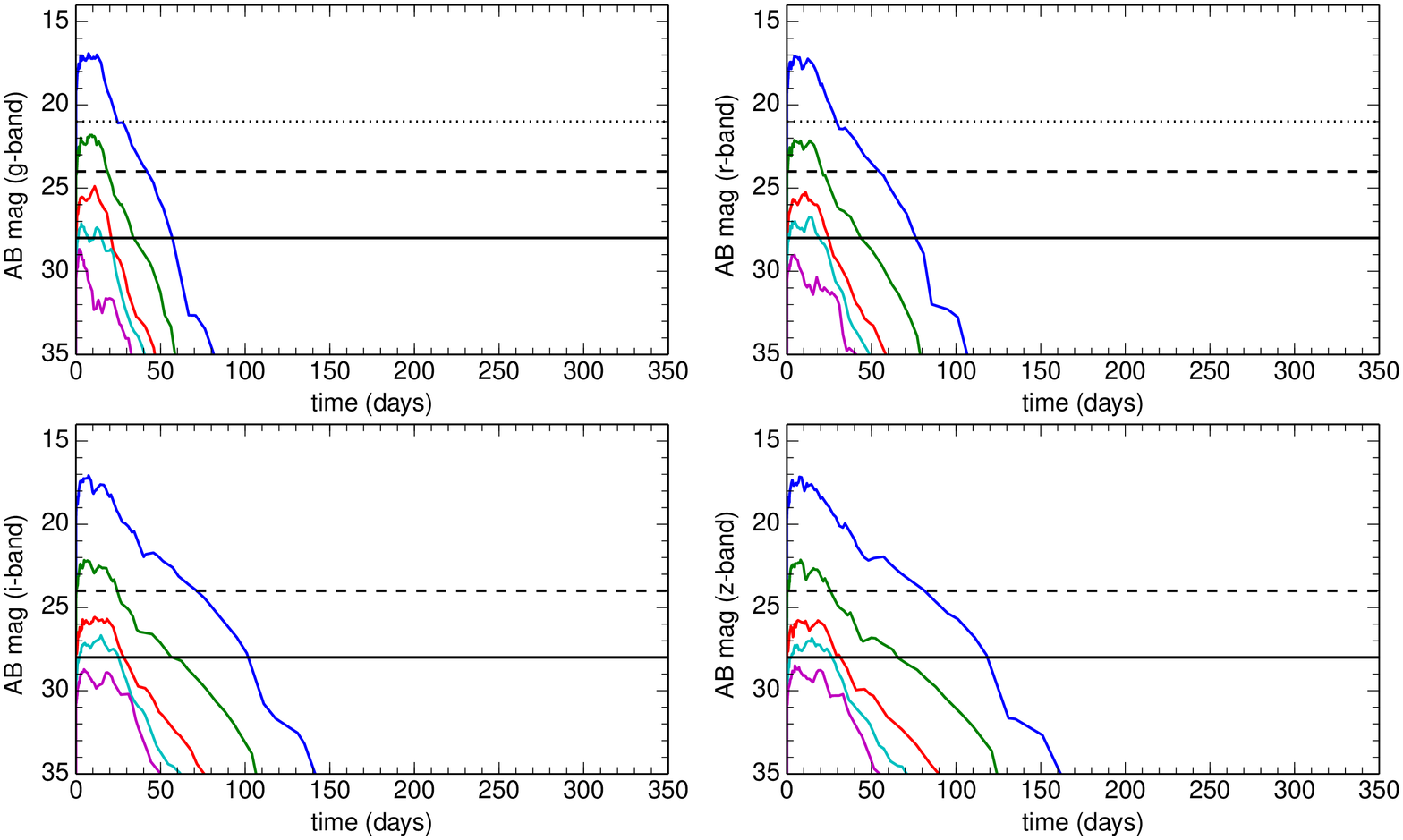,width=0.99\linewidth,clip=} \\
\epsfig{file=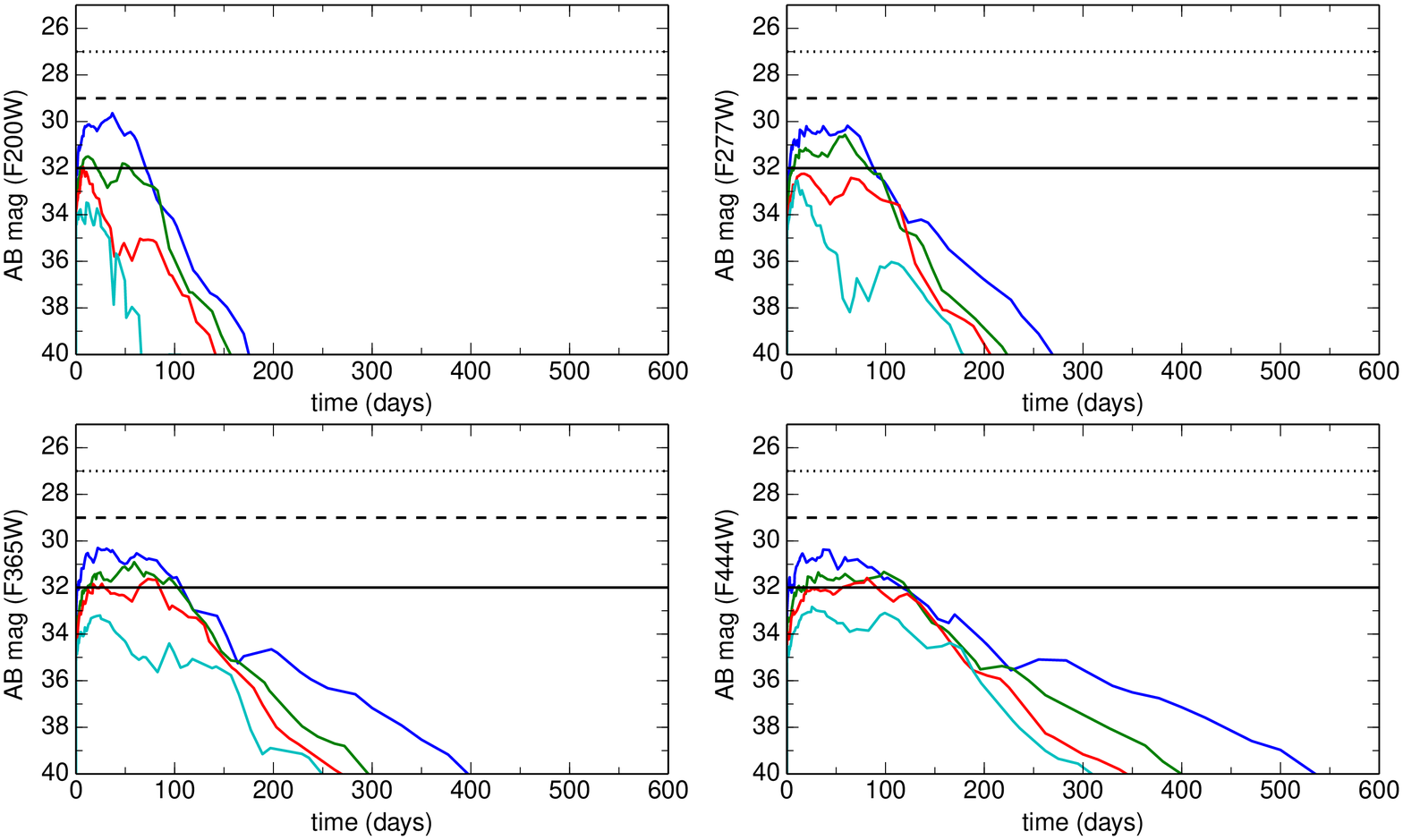,width=0.99\linewidth,clip=}
\end{tabular}
\end{center}
\caption{Light curves for the 52 foe 25 \Ms\ HN at low redshifts (upper panels) and 
high redshifts (upper panels).  In the upper panels, $z =$ 0.01 ({\it dark blue}), 0.1 
({\it green}), 0.5 ({\it red}), 1 ({\it light blue}), and 2 ({\it purple}).  The horizontal 
dotted, dashed and solid lines are detection limits for PTF, Pan-STARRS and LSST, 
respectively. In the lower panels, $z =$ 4 ({\it dark blue}), 7 ({\it green}), 10 ({\it red}), 
15 ({\it light blue}) and 20 ({\it purple}). The horizontal dotted, dashed and solid lines 
are detection limits for WFIRST, WFIRST with spectrum stacking and {\it JWST}, 
respectively.  The wavelength of each filter can be read from its name; for example, 
the F277W filter is centered at 2.77 $\mu$m, and so forth.}
\label{fig:52Bmag}
\end{figure*}

\begin{figure*}
\begin{center}
\begin{tabular}{c}
\epsfig{file=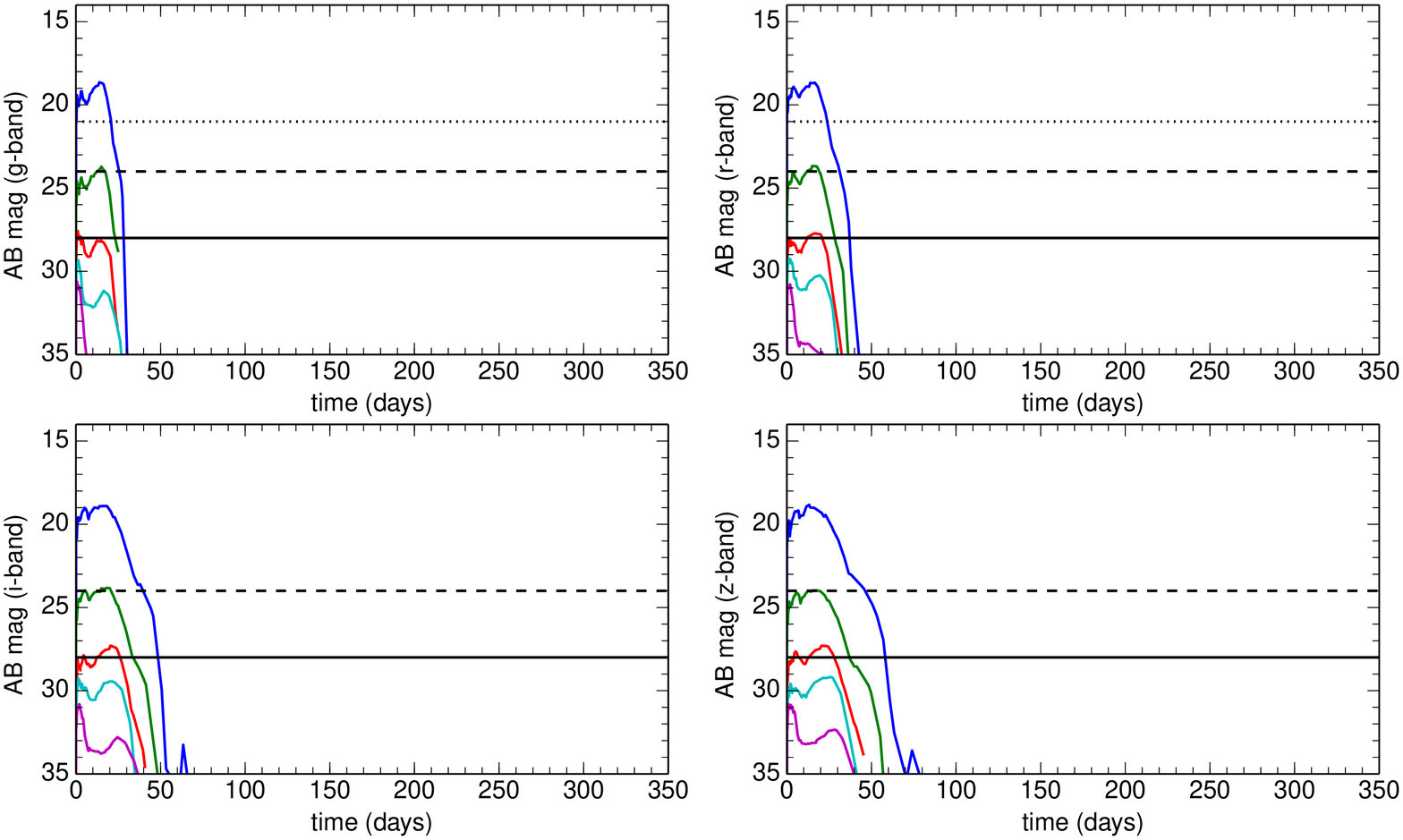,width=0.99\linewidth,clip=} \\
\epsfig{file=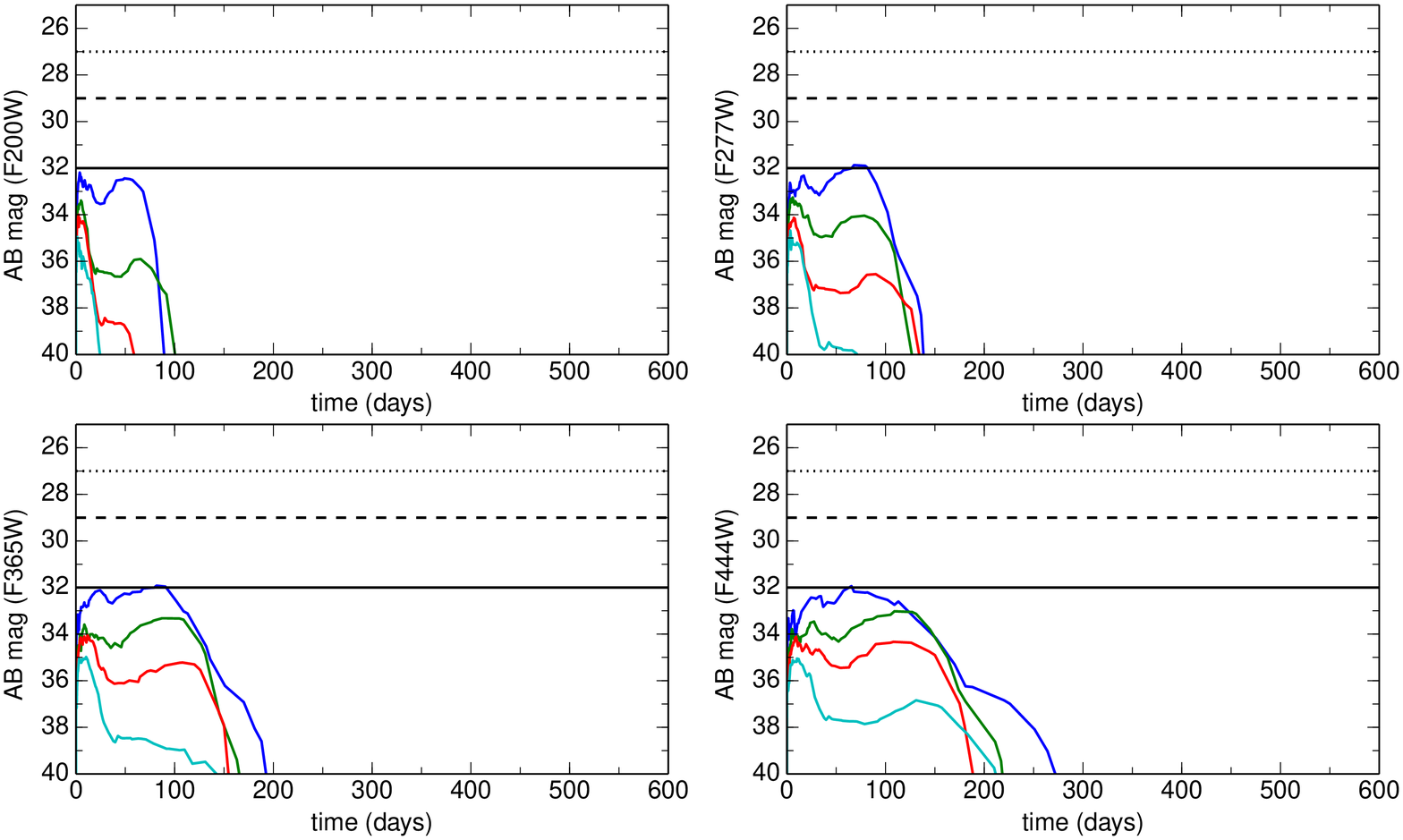,width=0.99\linewidth,clip=}
\end{tabular}
\end{center}
\caption{Light curves for the 22 foe 25 \Ms\ HN at low redshifts (upper panels) and 
high redshifts (upper panels).  In the upper panels, $z =$ 0.01 ({\it dark blue}), 0.1 
({\it green}), 0.5 ({\it red}), 1 ({\it light blue}), and 2 ({\it purple}).  The horizontal 
dotted, dashed and solid lines are detection limits for PTF, Pan-STARRS and LSST, 
respectively. In the lower panels, $z =$ 4 ({\it dark blue}), 7 ({\it green}), 10 ({\it red}), 
15 ({\it light blue}) and 20 ({\it purple}). The horizontal dotted, dashed and solid lines 
are detection limits for WFIRST, WFIRST with spectrum stacking and {\it JWST}, 
respectively.  The wavelength of each filter can be read from its name; for example, 
the F277W filter is centered at 2.77 $\mu$m, and so forth.}
\label{fig:22Bmag}
\end{figure*}

\begin{figure*}
\begin{center}
\begin{tabular}{c}
\epsfig{file=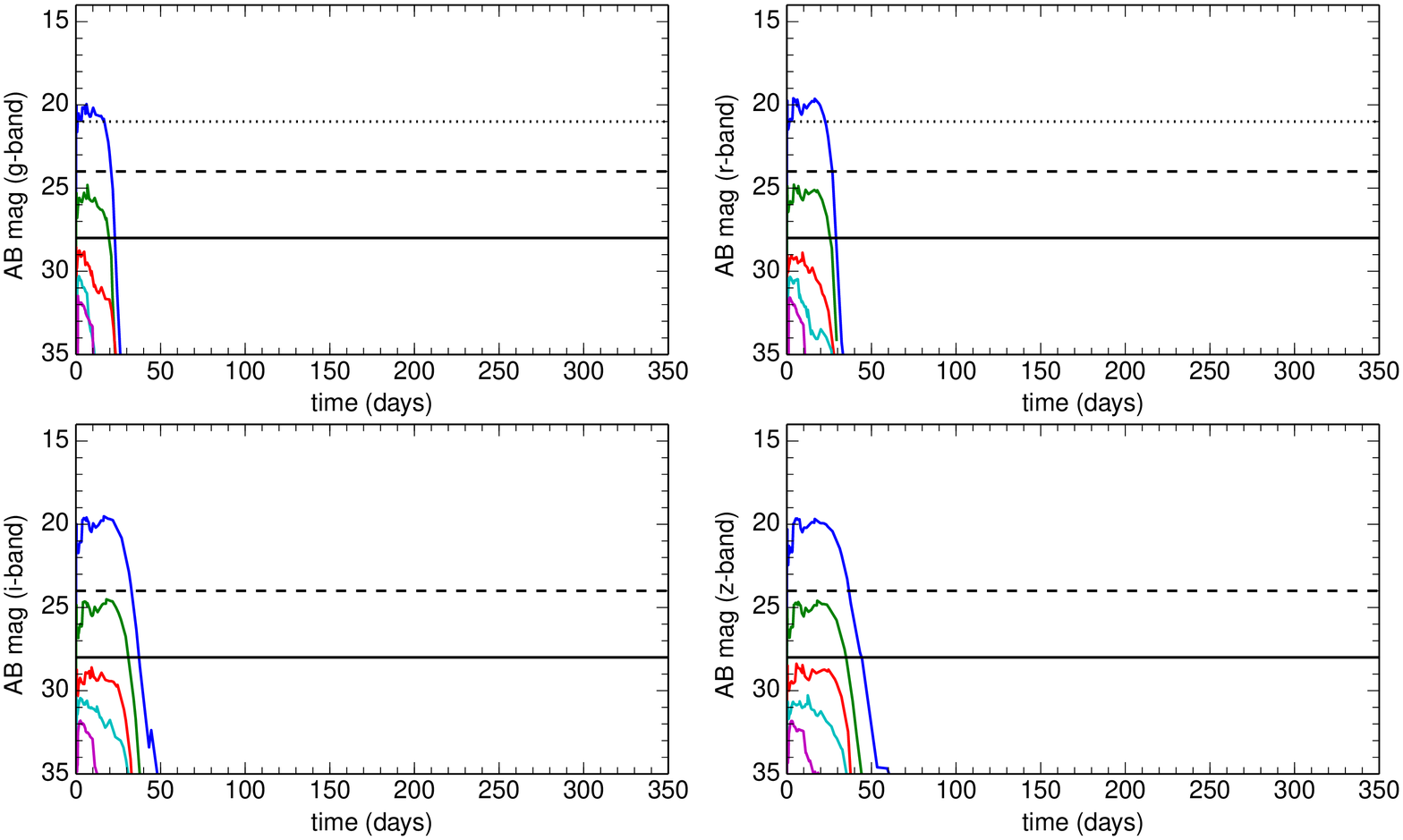,width=0.99\linewidth,clip=} \\
\epsfig{file=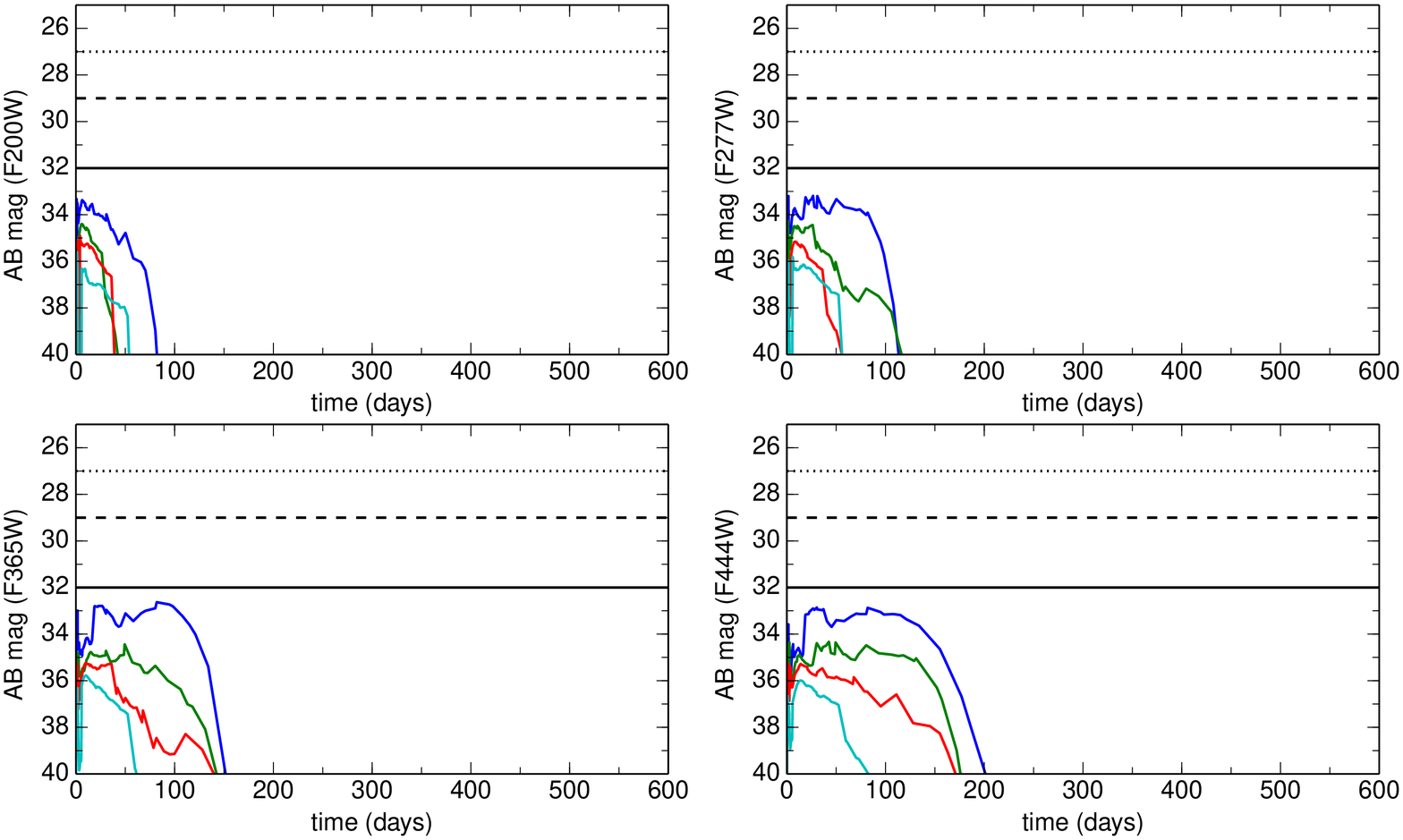,width=0.99\linewidth,clip=}
\end{tabular}
\end{center}
\caption{Light curves for the 10 foe 25 \Ms\ HN at low redshifts (upper panels) and 
high redshifts (upper panels).  In the upper panels, $z =$ 0.01 ({\it dark blue}), 0.1 
({\it green}), 0.5 ({\it red}), 1 ({\it light blue}), and 2 ({\it purple}).  The horizontal 
dotted, dashed and solid lines are detection limits for PTF, Pan-STARRS and LSST, 
respectively. In the lower panels, $z =$ 4 ({\it dark blue}), 7 ({\it green}), 10 ({\it red}), 
15 ({\it light blue}) and 20 ({\it purple}). The horizontal dotted, dashed and solid lines 
are detection limits for WFIRST, WFIRST with spectrum stacking and {\it JWST}, 
respectively.  The wavelength of each filter can be read from its name; for example, 
the F277W filter is centered at 2.77 $\mu$m, and so forth.}
\label{fig:10Bmag}
\end{figure*}

We show bolometric luminosities for all six HNe in Figure~\ref{fig:bolo} and hydro 
profiles for the 52 foe 50 \Ms\ HN in Figure~\ref{fig:hydro}.  We first examine shock
breakout from the star, as shown in the left column of Figure~\ref{fig:hydro}.  After 
breakout, the radiation pulse from the shock blows the outer layers of the star 
outward at $\sim$ 2.4 $\times$ 10$^{10}$ cm s$^{-1}$ as it descends the density 
bridge.  This radiative precursor stops accelerating as it reaches the bottom of the 
bridge.  Radiation breakout coincides with shock breakout.  The radiation front (the 
temperature plateau at 373 and 392 seconds) initially heats the gas to $\sim$ 200 
eV.  As the fireball expands, it cools by emitting photons and performing $PdV$ 
work on the envelope.  As it cools, its spectrum softens, and the temperature to 
which the radiation front heats the surrounding gas also falls.

Naively, one might expect the duration of the breakout transient to be roughly the light 
crossing time of the star.  It is actually longer in part because photons remain partially 
coupled to the wispy outer layers of the star that are blown off by the breakout pulse.  
As they diffuse out through this radiative precursor, they broaden the transient.  Also,
the opacities are frequency dependent, and photons break free of the flow at different
times at different wavelengths.  This effect also broadens the pulse in time \citep{bay14}.  
A few seconds after the precursor is blown off from the shock, at $\sim$ 390 seconds, 
photons escape from its outer layers and become visible to an external observer.  As 
shown in Figure~\ref{fig:bolo}, the bolometric luminosity of the breakout transient varies 
from $\sim$ 10$^{46}$ to 10$^{47}$ erg s$^{-1}$ and increases with explosion energy 
for a given stellar mass.  The transient is dimmer in more massive SNe at a given 
energy because of the greater inertia of the ejecta.  

Shock breakout happens earlier in the less massive star at a given energy because of 
its smaller radius. It happens earlier at higher energies because the shock reaches the 
surface sooner.  The breakout transient is composed mostly of X-rays and hard UV. 
At $z \sim 20$ the transient would would last up to 1 - 2 days today, in principle making 
it much easier to detect at this redshift than in the local universe. But while it is the most 
luminous phase of the SN, shock breakout is least visible at high redshifts due to 
absorption by the neutral IGM. Whatever X-rays that are not absorbed would redshifted 
into the far UV and stopped by the outer layers of our Galaxy.

Radiation from the shock sustains the radiative precursor until $\sim$ 2500 seconds, as 
shown in the center column of Figure~\ref{fig:hydro}.   The precursor is visible as the 
slightly noisy ramp in density between the shock and the surrounding wind at 3 $\times$
10$^{12}$ cm at 508 seconds.  It is also visible in the break in the velocity peak at the 
same position and time.  The shock soon overtakes the precursor and merges with it 
because it dims as it expands and cools, so its radiative flux can no longer maintain it. 

The rebrightening in the 22 and 52 foe explosions at $\sim$ 5 $\times 10^5$ s to 
$10^7$ s is due to the decay of \Ni\ in the ejecta.  It is brighter with greater explosion 
energy at a given progenitor mass because more \Ni\ is synthesized, and it is absent 
in the least energetic 25 \Ms\ and 50 \Ms\ HNe because they make very little \Ni.  
Rebrightening happens sooner with the 25 \Ms\ progenitor because of the shorter
radiation diffusion timescales in the ejecta.  We note that the expansion of the flow is 
nearly homologous after 10$^4$ seconds except for internal expansion of the \Ni\ 
bubble relative to the surrounding ejecta due to decay heating.  All six HNe evolve 
through similar stages.

\section{NIR Light Curves / Detection Limits}

NIR observations are required to detect SNe before the end of reionization ($z \sim
$ 6) because flux blueward of the Lyman limit at higher redshifts is absorbed by the 
neutral IGM.  This also limits detections of such events in the optical to $z < $ 6. 
All-sky surveys are probably the best prospects for detecting large numbers of high
$z$ SNe because their large survey areas can compensate for low star formation 
rates (SFRs) at early epochs \citep[e.g., Fig.~3 of][]{wet13c}. But even 30-m class 
telescopes with narrow fields of view such as {\it JWST}, the {\it Giant Magellan 
Telescope} ({\it GMT}), the Thirty-Meter Telescope (TMT), and the European 
Extremely Large Telescope (E-ELT) are still expected to find appreciable numbers 
of Pop III SNe \citep{hum12}.  We now consider detection limits in redshift for our  
HNe in the NIR for explosions at $z >$ 6 and in the optical for events below this 
redshift. 

In Figures~\ref{fig:52mag}-\ref{fig:10Bmag}, we show visible and NIR light curves 
for all six HNe along with detection limits for current and proposed instruments for 
$z =$ 0.01 - 20.  They were obtained from the spectra by summing their 
luminosities over the appropriate bands and then cosmologically redshifting and 
dimming them.  Since the NIR light curves are all redward of the Lyman limit in the 
frame of the explosion, we take the transmission coefficient of the neutral IGM at 
$z \gtrsim$ 6 to be 1 \citep[see Figure~3 of][]{ds13}.  The detection limits for the 
Palomar Transient Factory (PTF), the Panoramic Survey Telescope \& Rapid 
Response System (Pan-STARRS) and the Large Synoptic Survey Telescope 
(LSST) are AB mag 21, 24 and 28, respectively.  Photometry limits for {\it JWST} 
and WFIRST are AB mag 32 and 27, respectively, which could be extended to 29 
for WFIRST with spectrum stacking.  Note that wavelengths can be extracted from 
the {\it JWST} filter names by dividing their numbers by 100, i.e., the F277W is for 
2.77 $\mu$m.

The NIR light curves of the most energetic 50 \Ms\ HNe exhibit a initial, short-lived 
peak corresponding to the post-breakout expansion and cooling of the fireball 
followed by a second brighter and much longer peak due to \Ni\ rebrightening.  {\it 
JWST} detections of the dimmest 50 \Ms\ HNe will be restricted to $z \lesssim$ 4, 
but the more energetic ones will be visible out to $z =$ 10 - 15. The most energetic 
25 \Ms\ Hn will only be visible to {\it JWST} out to $z \sim$ 4 - 7.  WFIRST will only 
observe the brightest HNe out to $z \sim$ 4 - 5.  While {\it JWST} could therefore 
detect Pop III HNe in primordial galaxies if it happened across one, WFIRST in 
principle could see many more of these events, but only out to the end of 
cosmological reionization.  The fact that these SNe rise above photometry limits in 
multiple filters at a given redshift makes it easier to identify them as transients.  

The 22 and 52 foe 50 \Ms\ explosions will be visible to PTF out to $z \sim$ 0.01 - 
0.1, to Pan-STARRS out to $z \sim$ 0.1 - 0.5, and to LSST out to $z \sim$ 1 - 2.  
The 10 foe HN will be visible to PTF out to $z \sim$ 0.01, to Pan-STARRS out to 
$z \sim$ 0.01 - 0.1, and to LSST out to $z \sim$ 0.1 - 0.5.   The 25 \Ms\ HNe as
a rule are significantly dimmer and less long-lived than the 50 \Ms\ HNe.  The 25 
\Ms\ explosions will be visible to PTF out to $z \sim$ 0.01.  Pan-STARRS will 
detect these events out to $z \sim$ 0.01 - 0.1.  LSST will observe these HNe out 
to $z \sim$ 0.5 - 1, with most events being limited to $z \lesssim$ 0.5.    

\section{Pop III HN Rates}

Pop III HN rates are uncertain because the primordial IMF and star formation rates 
are unknown.  But there is reason to believe that Pop III stars end their lives as HNe 
at high enough rates to be detected in future surveys.  Today, the observed HN rate 
is within a factor of a few of the GRB rate, which is not surprising given that both are 
associated with Type Ib/c SNe whose progenitors are probably rapidly-rotating stars 
with masses $\ga$ 30 \Ms\ \citep{pd04,gdv07}.  Most Pop III stars are at least this 
massive, and many of them may be born with high rotation rates \citep{stacy11b,
stacy13,mm12} and die as GRBs or HNe more often than do stars of similar mass 
today \citep[e.g.,][]{yoon05,hmm05,wh06,nsi12,yoon12}).  There is also evidence 
that rapidly-rotating Pop III stars \citep{chiap11} and perhaps HNe \citep{nom10} 
may have synthesized the heavy elements detected in metal-poor stars, which also 
suggests that HNe may have been common among early stars.  

If Pop III HN rates also trace the GRB rate, several recent estimates of the Pop III 
GRB rate can be used to derive the Pop III HN rate.  \citet{bl06a} predict a total rate 
of observed Pop III GRBs of $\sim$ 0.1 yr$^{-1}$ at $z \gtrsim$ 15, which would 
correspond to $\ga$ 10 HNe yr$^{-1}$ if they can be seen at any viewing angle (i.e., 
if the beaming factor used to calculate the GRB rate is divided out).  Distinguishing 
between Pop III stars that are formed with and without radiative feedback during 
early galaxy formation, \citet{ds11} predict that GRBs from the former are orders of 
magnitude more common than those from the latter.  They predict an intrinsic GRB 
rate of $\ga$ 100 yr$^{-1}$ out to $z$ $\sim$ 15. \citet{camp11} use cosmological 
simulations of metal enrichment and Pop III and II star formation to place an upper 
limit of $\sim$ 1 yr$^{-1}$ on the observed Pop III GRB rate at $z$ $>$ 6, or about 
100 HNe yr$^{-1}$.

Even if no connection is assumed for Pop III GRBs and HNe, an upper limit to the 
HN rate can be gleaned from cosmological simulations of Pop III star formation that 
include chemical and radiative feedback \citep{jdk12} by assuming that all 25 - 140 
\Ms\ Pop III stars can produce black holes and HNe \citep[e.g.,][]{fryer99,het03}.  
Assuming a Salpeter-like IMF for primordial stars and a lower mass limit of 21 \Ms, 
\citet{jdk12} find a total HN rate of $\sim$ 10$^4$ yr$^{-1}$, with most occurring at 
$z$ $\la$ 10. This rate is broadly consistent with the estimates above if the actual 
ratio of Pop III black hole-producing SNe to Pop III HNe is similar to the observed 
ratio today \citep[$\sim$ 100;][]{pd04}.  We conclude that HNe could occur at rates 
of $\ga$ 10 yr$^{-1}$, and perhaps up to $\ga$ 100 yr$^{-1}$, at $z$ $\la$ 15.

\section{Conclusion}

Pop III HNe will be visible in the NIR out to $z \sim$ 10 - 15 by {\it JWST} and out 
to $z \sim$ 4 - 5 to WFIRST and WISH.  These redshifts would go up dramatically 
if the HN crashes into a massive shell ejected by the progenitor.  Such collisions 
can produce superluminous supernovae (SLSNe) like SN 2006gy \citep{nsmith07b,
moriya10,chev11,moriya12} that would be brighter than the HN itself.  The high 
luminosities of SLSNe are due to the large radius of the shell upon impact.  Much 
less energetic Pop III Type IIn SNe (2 foe) can be detected by {\it JWST} at $z \sim
$ 15 - 20 and WFIRST at $z \sim$ 7 \citep{wet12e}, so HNe that collide with dense 
shells may be visible to all-sky NIR missions like WFIRST out to $z \sim$ 10 - 15. 
This would greatly enhance their prospects for detection, since the large survey 
areas of these missions could compensate for low HN rates. We are now modeling 
these events with RAGE.  

The ejection of the H layer prior to the HN could create a dense wind rather than 
a shell, and this envelope could also affect the luminosity of the explosion \citep{
bay14}.  When the shock crashes into this wind it could become even hotter, and 
thus more luminous.  But more of this luminosity may also be absorbed by the 
envelope downwind of the shock. Additional simulations are required to determine 
the overall effect of dense envelopes on the luminosity of the HN.  We considered 
only HNe in very diffuse winds, in which all vestiges of the H layer are driven 
beyond the immediate reach of the ejecta, as a simplest case.

Many of the HNe detected by {\it JWST} at $z \sim$ 10 - 15 could be zero-metallicity
events, especially in cases where supersonic baryonic streaming motion delays first 
star formation to $z \sim$ 15 - 17 \citep{th10,greif11}. But HNe found by WFIRST at
$z \sim$ 4 - 5 would probably not be Pop III explosions, even though \citet{tss09} 
have found that Pop III star formation could extend down to $z \sim$ 6, and large 
pockets of metal-free gas have now been discovered at $z \sim$ 2 \citep{fop11}. 
How might HNe at $Z \sim$ 0.1 \Zs\ differ from Pop III events?  It has been found 
that the central engines of core-collapse explosions do not vary strongly with 
metallicity because the cores of $Z =$ 0 and \Zs\ stars have similar entropy profiles 
\citep{cl04,wh07} \citep[see also Figure~1 of][]{wf12}.  It is therefore likely that HNe 
at the end of reionization would have similar energies to those in the primordial 
universe.  

Although we have modeled Pop III HNe with 1D simulations, they are inherently
multidimensional events because of their highly asymmetric central engines.  Real 
HNe may therefore exhibit azimuthally-dependent luminosities that could affect not 
only their detection limits in the NIR at high redshift but how many of these events 
would actually be seen for a given opening angle for the engine.  Future 2D radiation 
hydrodynamical simulations could address these issues.Ê But for now, as with other 
studies \citep{moriya12}, we must rely on 1D models to estimate the NIR signatures 
of these events.  Because our simulations are 1D, they also neglect mixing during 
the explosion.  Mixing could have some impact on the luminosity if it dredges up \Ni\ 
from greater depths during the explosion.  Mixing can also affect the order in which 
lines appear in the spectra over time.

If most Pop III HNe are associated with GRBs, and most GRBs are due to binary 
mergers with companion stars \citep[e.g.,][]{fw98,fwh99,zf01,pasp07}, then there 
is additional reason to believe HNe may have been common in the primordial 
universe because Pop III stars have now been found to form in binaries and small
multiples in simulations \citep{turk09,stacy10}.  The GRBs themselves might be 
detected by other means.  Gamma rays from these events could trigger {\it Swift} 
or its successors, such as the Joint Astrophysics Nascent Universe Satellite 
\citep[JANUS,][]{mesz10,Roming08,burrows10}, and their afterglows \citep{wet08c} 
might be found in all-sky radio surveys by the Extended Very Large Array (eVLA), 
eMERLIN and the Square Kilometer Array (SKA) \citep{ds11} \citep[see also][]{
suwa11,nsi12}.  It is now known that Pop III GRB afterglows will be bright enough 
in the NIR to be seen {\it JWST}, WFIRST, and the TMT \citep{met12a,mes13a}
(and that they would completely outshine the HN).

Could later stages of HNe be detected in other ways?  \citet{wet08a} found that 
most of the kinetic energy of 40 \Ms\ Pop III HNe is eventually radiated away as 
H and He lines in primordial halos as the remnant sweeps up and shocks gas. 
This emission is too diffuse, redshifted and extended over time to be detected 
by any upcoming instruments.  Also, unlike Pop III PI SNe, HNe do not inject 
enough energy into the cosmic microwave background (CMB) to impose excess 
power on the CMB at small scales \citep{oh03,wet08a} or be directly imaged by 
the Atacama Cosmology Telescope or the South Pole Telescope via the 
Sunyaev-Zeldovich effect.  But new calculations reveal that enough synchrotron 
emission from their remnants would redshifted into the radio above $z \sim$ 10 
to be directly detected by current facilities such as eVLA and eMERLIN and by 
SKA \citep{mw12}.  Whether in the NIR, radio, or in the fossil abundance record, 
these ancient explosions could soon open another window on the $z \sim$ 10 - 
15 universe.

\acknowledgments

JS and JLJ were supported by LANL LDRD Director's Fellowships.  D.J.W. was 
supported by the European Research Council under the European Community's 
Seventh Framework Programme (FP7/2007 - 2013) via the ERC Advanced Grant 
"STARLIGHT:  Formation of the First Stars" (project number 339177).  Work at 
LANL was done under the auspices of the National Nuclear Security 
Administration of the U.S. Department of Energy at Los Alamos National 
Laboratory under Contract No. DE-AC52-06NA25396. All CASTRO, RAGE and 
SPECTRUM calculations were performed on Institutional Computing (IC) and 
Yellow network platforms at LANL (Pinto, Mustang and Moonlight).

\bibliographystyle{apj}
\bibliography{refs}

\end{document}